\documentclass[twocolumn]{jpsj3}
\usepackage{txfonts}
\usepackage{color}

\def\gsim{\mathop {\vtop {\ialign {##\crcr 
$\hfil \displaystyle {>}\hfil $\crcr \noalign {\kern1pt \nointerlineskip } 
$\,\sim$ \crcr \noalign {\kern1pt}}}}\limits}
\def\lsim{\mathop {\vtop {\ialign {##\crcr 
$\hfil \displaystyle {<}\hfil $\crcr \noalign {\kern1pt \nointerlineskip } 
$\,\,\sim$ \crcr \noalign {\kern1pt}}}}\limits}

\title{Basic Properties of Conductivity and Normal Hall Effect in the Periodic Anderson Model}

\author{\name{Shinji \name{Watanabe}}$^1$ and \name{Kazumasa \name{Miyake}}$^2$}
\inst{
$^1$Department of Basic Sciences, Kyushu Institute of Technology, Kitakyushu, Fukuoka 804-8550, Japan \\
$^2$Toyota Physical and Chemical Research Institute, Nagakute, Aichi 480-1192, Japan \\
} 

\abst{
Exact formulas of diagonal conductivity $\sigma_{xx}$ and Hall conductivity $\sigma_{xy}$ are derived from the Kubo formula in hybridized two-orbital systems with arbitrary band dispersions. On the basis of the theoretical framework for the Fermi liquid based on these formulas, the ground-state properties of the periodic Anderson model with electron correlation and weak impurity scattering are studied on the square lattice. It is shown that imbalance of the mass-renormalization factors in $\sigma_{xx}$ and $\sigma_{xy}$ causes remarkable increase in the valence-fluctuation regime as the f level increases while the cancellation of the renormalization factors causes slight increase in $\sigma_{xx}$ and $\sigma_{xy}$ in the Kondo regime. The Hall coefficient $R_{\rm H}$ shows almost constant behavior in both the regimes. Near half filling, $R_{\rm H}$ is expressed by the total hole density as $R_{\rm H}=1/(\bar{n}_{\rm hole}e)$ while $R_{\rm H}$ approaches zero near quarter filling, which reflects the curvature of the Fermi surface. These results hold as far as the damping rate for f electrons is less than about $10~\%$ of the renormalized hybridization gap. 
From these results we discuss pressure dependence of residual resistivity and normal Hall effect in Ce- and Yb-based heavy electron systems.
}

\begin{document}
\maketitle 

\section{Introduction}

The heavy-electron systems have attracted much attention in condensed matter physics since the localized and itinerant natures of strongly correlated electrons open up unexpected phenomena, which leads to a break through the paradigm to provide a new concept with universality. 
The emergence of the heavy mass of quasiparticles, their condensation to the superconductivity with essentially ``high" transition temperature compared to the renormalized Fermi temperature, and unconventional quantum critical phenomena~\cite{MW2014,WM2010} are such examples. 

To make  experimental explorations, the measurements of the conductivity and Hall effect have been performed extensively in the heavy-electron systems. 
Accumulated data for the Ce-based heavy electron systems show that there exists a general tendency that the isothermal resistivity at the measured lowest temperature, i.e., residual resistivity, in the Fermi-liquid regime decreases as pressure increases~\cite{Kagayama1991,Kagayama1996,Raymond}. On the other hand, in the Yb-based heavy-electron systems, there exits a general tendency that the residual resistivity in the Fermi-liquid regime increases as pressure increases~\cite{Jaccard1998,Knebel2001,Yuan2006,Matsubayashi,Honda2010,Kim2013}.  
The mechanism of the pressure dependence of the conductivity and also the Hall conductivity 
in relation to the valence of Ce and Yb as well as the shape of the Fermi surface has been desired to be clarified theoretically. 

Theoretically, 
the diagonal conductivity and the Hall conductivity were formulated on the basis of the Boltzmann transport 
theory~\cite{Ziman,Tsuji,Ong}. 
The formula of the diagonal conductivity in the system with Coulomb repulsion among electrons 
was derived microscopically by \'{E}liashberg~\cite{Eliashberg} in the Fermi liquid theory 
starting from the Kubo formula~\cite{Kubo}.
The diagonal conductivity in the periodic Anderson model 
which is the prototypical model for the Ce- and Yb-based heavy electron systems 
was formulated  
on the basis of the Fermi liquid theory by Yamada and Yosida~\cite{Yamada}. 

The Hall conductivity was formulated by Fukuyama~\cite{FEW,Fukuyama} in the gauge invariant manner 
starting from the Kubo formula. 
Later, 
\'{E}liashberg's work was extended to the Hall conductivity by Kohno and Yamada, 
who formulated the general expression in the Femi liquid theory~\cite{Kohno}. 
Explicit calculation of 
the diagonal and Hall conductivities was performed  
by using the conserving approximation~\cite{BK,Baym,Bickers} 
in the Hubbard model which contains a single orbital with electron transfer and Coulomb repulsion by Kontani {\it et al}~\cite{Kontani1999}. 

So far, after the formulation of the diagonal conductivity by Yamada and Yosida, systematic calculations for the diagonal and normal Hall conductivities have not been reported 
in detail 
in the periodic Anderson model which contains the two orbitals for f and conduction electrons. 

In this paper, we clarify the basic properties of the  diagonal and Hall conductivities, and normal Hall effect in the periodic Anderson model. On the basis of the theoretical framework which describes the Fermi liquid correctly, the ground-state properties of the diagonal and Hall conductivities and the Hall coefficient are studied by taking into account the effects of the electron correlation and the weak impurity scattering. By performing the numerical calculation on the square lattice, the dependence on parameters such as the f level, c-f hybridization, the damping rate for f electrons, and the filling is clarified. The relation to the shape of the Fermi surface and the f-electron number per site, corresponding to the valence of Ce and Yb, is also clarified.   

The organization of this paper is as follows: 
In Sect.~\ref {sec:PAM_finiteU}, we review the formalism of conductivity in the periodic Anderson model based on the Fermi-liquid theory. In Sect.~\ref{chap:U0}, exact formulas of the diagonal conductivity and Hall conductivity are derived in the periodic Anderson model 
for $U=0$. In Sect.~\ref{chap:SL}, the ground-state properties of the conductivities and the Hall coefficient in the periodic Anderson model with electron correlations and weak-impurity scattering are studied on the basis of the exactly derived formulas. The paper is summarized in Sect.~5.

We take the energy units as $\hbar=1$, $k_{\rm B}=1$, and the light velocity $c=1$. 
Note that we denote $e$ as the elementary charge, 
i.e., $e>0$. 

\section{Formalism of conductivity based on the Fermi liquid theory 
in the periodic Anderson model 
}
\label{sec:PAM_finiteU}
As the simplest minimal model for the electronic state in the Ce- and Yb-based heavy-electron systems, we consider the periodic Anderson model
\begin{eqnarray}
{\cal H}&=&
\sum_{{\bf k}\sigma}\varepsilon_{\bf k}
c_{{\bf k}\sigma}^{\dagger}c_{{\bf k}\sigma}
+\sum_{{\bf k}\sigma}\varepsilon^{\rm f}_{\bf k}f_{{\bf k}\sigma}^{\dagger}f_{{\bf k}\sigma}
+\sum_{{\bf k}\sigma}V_{\bf k}
\left(f_{{\bf k}\sigma}^{\dagger}c_{{\bf k}\sigma}+c_{{\bf k}\sigma}^{\dagger}f_{{\bf k}\sigma}
\right)
\nonumber
\\
&+&U\sum_{i}n_{i\uparrow}^{\rm f}n_{i\downarrow}^{\rm f},
\label{eq:PAM}
\end{eqnarray}
with $n_{i\sigma}^{\rm f}\equiv f_{i\sigma}^{\dagger}f_{i\sigma}$ 
where  $f_{i\sigma}$ $(f_{i\sigma}^{\dagger})$ is an annihilation (creation) operator 
of an f electron at the $i$-th site with spin $\sigma$ and 
$c_{{\bf k}\sigma}$ $(c_{{\bf k}\sigma}^{\dagger})$ is an annihilation (creation) operator 
of a conduction electron at the wave vector $\bf k$ with spin $\sigma$. 
The first term represents the energy band of conduction electrons with a dispersion, $\varepsilon_{\bf k}$. 
The second term represents the energy of f electrons allowed to have a dispersion, 
$\varepsilon^{\rm f}_{\bf k}$, which we here consider for generality. 
The third term represents the hybridization between f and conduction electrons 
with real $V_{\bf k}$. 
The on-site Coulomb repulsion for f electrons is expressed in the last term. 

Since 4f$^1$5d$^1$6s$^2$ configuration is realized in the outermost electrons' shell of Ce, Ce$^{+3}$ contains the 4f$^1$ electron and Ce$^{+4}$ contains the 4f$^0$ electron. On the other hand, since 4f$^{14}$6s$^2$ configuration is realized in the outermost electrons' shell of Yb, Yb$^{3+}$ contains the 4f$^{13}$ electrons and Yb$^{2+}$ contains 4f$^{14}$ electrons. Since 4f$^{14}$ is the closed shell for the f orbital, by taking the hole picture instead of the electron picture, Eq.~(\ref{eq:PAM}) can be applied to the Yb-based systems and the parallel discussion to the Ce-based systems in the electron picture can be made.   

The total filling $\bar{n}$ is defined as 
\begin{eqnarray}
\bar{n}\equiv n_{\rm f}+n_{\rm c},
\label{eq:filling_PAM}
\end{eqnarray}
where $n_{\rm f}$ and $n_{\rm c}$ are f-electron number per site 
and the conduction electron number per site, respectively, defined as 
\begin{eqnarray}
n_{\rm f}&=&\frac{1}{N}\sum_{i\sigma}\langle n_{i\sigma}^{\rm f}\rangle,
\\
n_{\rm c}&=&\frac{1}{N}\sum_{i\sigma}\langle n_{i\sigma}^{\rm c}\rangle. 
\end{eqnarray}
Here, $N$ is the number of the lattice sites and $n_{i\sigma}^{\rm c}\equiv c_{i\sigma}^{\dagger}c_{i\sigma}$. 
Note that $\bar{n}=2$ is the half filling.

The diagonal conductivity in the periodic Anderson model was formulated 
on the basis of the Fermi liquid theory  
by Yamada and Yosida~\cite{Yamada}.  
In the following, we review the formalism for the diagonal conductivity. 

The retarded Green functions of f electrons and conduction electrons are given by
\begin{eqnarray}
G_{\bf k}^{\rm ff \ R}(\varepsilon)
&=&\left[
\varepsilon+i\delta-\varepsilon^{\rm f}_{\bf k}-\Sigma_{\bf k}^{\rm R}(\varepsilon)
-\frac{V_{\bf k}^2}{\varepsilon+i\delta-\varepsilon_{\bf k}}
\right]^{-1},
\\
G_{\bf k}^{\rm cc \ R}(\varepsilon)
&=&\left[
\varepsilon+i\delta-\varepsilon_{\bf k}-\frac{V_{\bf k}^2}
{\varepsilon+i\delta-\varepsilon^{\rm f}_{\bf k}-\Sigma_{\bf k}^{\rm R}(\varepsilon)}
\right]^{-1}, 
\end{eqnarray}
respectively, 
where $\Sigma_{\bf k}^{\rm R}(\varepsilon)$ is the retarded self energy of f electrons, 
which arises from the Coulomb repulsion $U$, 
and $\delta$ is the infinitesimal positive constant. 

We consider the case where the Fermi level is located at the lower hybridized band. 
In the vicinity of the Fermi energy, the Green functions of f and conduction electrons 
are described by quasiparticles as  
\begin{eqnarray}
G_{\bf k}^{\rm ff R}(\varepsilon)&=&a_{-,{\bf k}}^{\rm ff}
G_{\bf k}^{\rm ff  -R}(\varepsilon),
\label{eq:Gff_U} 
\\
G_{\bf k}^{\rm cc R}(\varepsilon)&=&a_{-,{\bf k}}^{\rm cc}
G_{\bf k}^{\rm ff  -R}(\varepsilon), 
\end{eqnarray}
respectively, 
where 
the retarded Green function for the lower hybridized band is given by
\begin{eqnarray}
G_{\bf k}^{\rm ff  -R}(\varepsilon)=\frac{1}{\varepsilon-E^{-*}_{\bf k}+i\Gamma_{\bf k}^{*}}
\end{eqnarray}
and 
the renormalization factors are given by
\begin{eqnarray}
a_{-,{\bf k}}^{\rm ff}&=&
\left.
\left[
1-\frac{\partial{\rm Re}\Sigma_{\bf k}^{\rm R}(\varepsilon)}{\partial\varepsilon}
+\frac{V_{\bf k}^2}{(\varepsilon-\varepsilon_{\bf k})^2}
\right]^{-1}
\right|_{\varepsilon=E_{\bf k}^{-*}},
\label{eq:aff_U}
\\
a_{-,{\bf k}}^{\rm cc}&=&
\left(
\frac{V_{\bf k}}{E_{\bf k}^{-*}-\varepsilon_{\bf k}}
\right)^2
a_{-,{\bf k}}^{\rm ff}, 
\label{eq:acc_U}
\end{eqnarray}
respectively. 
Here, the renormalized lower hybridized band is given by
\begin{eqnarray}
E^{-*}_{\bf k}=\frac{\varepsilon_{\bf k}+\tilde{\varepsilon}_{\bf k}^{\rm f}}{2}
-\frac{1}{2}\sqrt{\left(\varepsilon_{\bf k}-\tilde{\varepsilon}_{\bf k}^{\rm f}\right)^2+4{\tilde{V}_{\bf k}}^2}
\end{eqnarray}
with the renormalized f level 
\begin{eqnarray}
\tilde{\varepsilon}_{\bf k}^{\rm f}\equiv 
z_{\bf k}
\left[\varepsilon_{\bf k}^{\rm f}
+{\rm Re}\Sigma_{{\bf k}}^{\rm R}(\mu)
\right], 
\end{eqnarray}
and the renormalized hybridization 
${\tilde{V}_{\bf k}}^2\equiv z_{\bf k} V_{\bf k}^2$. 
Here, $z_{\bf k}$ is defined as
\begin{eqnarray}
z_{\bf k}\equiv
\left.
\left[
1-\frac{\partial{\rm Re}\Sigma_{\bf k}^{\rm R}(\varepsilon)}{\partial\varepsilon} 
\right]^{-1}\right|_{\varepsilon=\mu}, 
\end{eqnarray}
and $\mu$ is the chemical potential. 
The renormalized damping rate is given by
\begin{eqnarray}
\Gamma_{\bf k}^{*}=a_{-,{\bf k}}^{\rm ff}\left(-{\rm Im}\Sigma_{\bf k}^{\rm R}(\mu)\right)>0. 
\label{eq:Gamma_renm}
\end{eqnarray}
The diagonal conductivity is expressed as
\begin{eqnarray}
\sigma_{xx}&=&
\frac{e^2}{V_0}
\sum_{\bf k}\int_{-\infty}^{\infty}\frac{d\varepsilon}{\pi}
\left(
-\frac{\partial f(\varepsilon)}{\partial\varepsilon}
\right)
\left\{
\left|G_{\bf k}^{\rm ff \ R}(\varepsilon)\right|^2
v_{{\bf k}x}(\varepsilon)
J_{{\bf k}x}(\varepsilon)
\right.
\nonumber
\\
& &
\left.
-{\rm Re}\left[
{{G_{\bf k}^{\rm ff \ R}}^2(\varepsilon)}
v_{{\bf k}x}^2(\varepsilon)
\right]
\right\},
\label{eq:sxx_full}
\end{eqnarray}
where $f(\varepsilon)$ is the Fermi distribution function 
$f(\varepsilon)=[{\rm e}^{(\varepsilon-\mu)/T}+1]^{-1}$ and 
$G_{\bf k}^{\rm ff \ R}(\varepsilon)$ is the f-electron Green function for quasiparticles 
given by Eq.~(\ref{eq:Gff_U}). 
The second term in Eq.~(\ref{eq:sxx_full}) was not described in Ref.~\citen{Yamada},  
but this term is necessary quantitatively~\cite{Kontani1999}. 
For example, 
the exact formula of $\sigma_{xx}$ derived for $U=0$ in Sect.~\ref{chap:U0} 
is  correctly reproduced by Eq.~(\ref{eq:sxx_full}) 
with the second term in the brace 
when we set 
$\Sigma_{\bf k}^{\rm R}(\varepsilon)=0$ [see Eq.~(\ref{eq:sxx_Boltzmann})].
Here, $v_{{\bf k}x}(\varepsilon)$ is the total velocity defined as 
\begin{eqnarray}
v_{{\bf k}x}(\varepsilon)=v_{{\bf k}x}^{\rm f}(\varepsilon)
+\left(\frac{V_{\bf k}}{\varepsilon-\varepsilon_{\bf k}}
\right)^2v_{{\bf k}x}^{{\rm c}0}
+\frac{\frac{\partial V_{\bf k}^2}{\partial k_{x}}}{\varepsilon-\varepsilon_{\bf k}}, 
\label{eq:v_def}
\end{eqnarray}
where the velocity of f electrons and conduction electrons are given by 
\begin{eqnarray}
v_{{\bf k}x}^{\rm f}(\varepsilon)&=&
v_{{\bf k}x}^{\rm f0}
+\frac{\partial {\rm Re}\Sigma_{\bf k}^{\rm R}(\varepsilon)}{\partial k_x}, 
\\
v_{{\bf k}x}^{\rm f0}&=&\frac{\partial\varepsilon_{\bf k}^{\rm f}}{\partial k_x},
\\
v_{{\bf k}x}^{\rm c0}&=&\frac{\partial\varepsilon_{\bf k}}{\partial k_x}, 
\label{eq:v_def2}
\end{eqnarray}
respectively. 
The total current $J_{{\bf k}x}(\varepsilon)$ is given by
\begin{eqnarray}
J_{{\bf k}x}(\varepsilon)=v_{{\bf k}x}(\varepsilon)
+\frac{1}{V_0}\sum_{\bf k'}
\int_{-\infty}^{\infty}\frac{d\varepsilon'}{4\pi i}
{\cal T}_{\bf kk'}(\mu,\varepsilon')
\left|G_{\bf k'}^{\rm ff \ R}(\varepsilon')\right|^2 J_{{\bf k'}x}(\varepsilon'),
\label{eq:J}
\end{eqnarray}
where ${\cal T}_{\bf kk'}(\mu,\varepsilon')$ is the irreducible four-point vertex 
introduced by \'{E}liashberg~\cite{Eliashberg}. 
On the basis of the framework of the conserving approximation~\cite{BK,Baym}  
where the general vertex corrections for the total current 
are considered to be consistent with the self-energy corrections. 
Yamada and Yosida pointed out that 
the total current $J_{{\bf k}x}(\varepsilon)$ can be obtained as 
the correct solution of the Bethe-Salpeter equation~(\ref{eq:J}) because of the presence of 
the Umklapp processes on the periodic crystal lattice~\cite{Yamada}. 
In other words, they stressed that the conductivity due to electron interaction 
should diverge in the absence of 
the Umklapp processes as the result of the momentum conservation.

In the case of $\Gamma^{*}\ll T$, 
the $\varepsilon$ integration in the first-term in Eq.~(\ref{eq:sxx_full}), 
which is denoted by $\sigma_{xx}^{(1)}$,
can be performed as  
\begin{eqnarray}
\sigma_{xx}^{(1)}&{\approx}&
\frac{e^2}{V_{0}}\sum_{\bf k}
\left(-\frac{\partial f(E_{\bf k}^{-*})}{\partial E_{\bf k}^{-*}}\right)
a_{-,{\bf k}}^{\rm ff}v_{{\bf k}x}(\mu)
\frac{J_{{\bf k}x}(\mu)}{-{\rm Im}\Sigma_{\bf k}^{\rm R}(\mu)}, 
\\
&=&
\frac{e^2}{V_0}\sum_{\bf k}
\delta\left(
\mu-E_{\bf k}^{-*}
\right)
a_{-,{\bf k}}^{\rm ff}v_{{\bf k}x}(\mu)
\frac{J_{{\bf k}x}(\mu)}{-{\rm Im}\Sigma_{\bf k}^{\rm R}(\mu)},
\label{eq:sxx_cancel}
\end{eqnarray}
where the last equation is derived for sufficiently low temperatures. 
Since the quasiparticle band $E_{\bf k}^{-*}$ near ${\bf k}\approx{\bf k}_{\rm F}$ 
is renormalized by the factor $z_{\bf k}$, $\delta(\mu-E_{\bf k}^{-*})$ 
is enhanced by $z_{\bf k}^{-1}$. 
Namely, the mass-renormalization factors $z_{\bf k}^{-1}$ and $a_{-,{\bf k}}^{\rm ff}$ 
which includes $z_{\bf k}$ [see Eq.~(\ref{eq:aff_U})] 
cancel out each other in Eq.~(\ref{eq:sxx_cancel}). 
This implies that in $\sigma_{xx}$, all renormalizations cancel out 
and the resistivity is proportional to 
${-{\rm Im}\Sigma_{\bf k}(\mu)}$. 
Since within the Fermi liquid theory 
the imaginary part of the self energy is proportional to $T^2$ at low temperatures, 
${-{\rm Im}\Sigma_{\bf k}(\mu)}\propto T^2$,  
the resistivity $\rho_{xx}=1/\sigma_{xx}$ shows the $T^2$ dependence~\cite{Yamada}. 

As for the normal Hall conductivity, 
by applying the formalism by Kohno and Yamada~\cite{Kohno} to the periodic Anderson model 
in the case where the single-band treatment is justified with small $\Gamma^{*}_{\bf k}$ 
at sufficiently low temperatures, the Hall conductivity for the lower-hybridized band is given by 
\begin{eqnarray}
\sigma_{xy}/H&\approx&\frac{e^3}{V_0}\sum_{\bf k}\int_{-\infty}^{\infty}\frac{d\varepsilon}{\pi}
\left(
-\frac{\partial f(\varepsilon)}{\partial\varepsilon}
\right)
\left|G_{\bf k}^{\rm ff \ R}(\varepsilon)\right|^2
{\rm Im}\left[G_{\bf k}^{\rm ff \ R}(\varepsilon)\right]
\nonumber
\\
& &
\times
v_{{\bf k}x}
\left[
J_{{\bf k}x}\frac{\partial J_{{\bf k}y}}{\partial k_{y}}
-J_{{\bf k}y}\frac{\partial J_{{\bf k}x}}{\partial k_{y}}
\right],    
\label{eq:sxy_full}
\end{eqnarray}
where $H$ is a weak magnetic field applied along the $z$ axis. 
Here, $J_{{\bf k}y}$ is the $y$ component of the total current vector ${\bf J}_{\bf k}$, 
which is given by setting $y$ instead of $x$ in Eq.~(\ref{eq:J}). 

In the case of $\Gamma_{\bf k}^{*}\ll T$, the $\varepsilon$ integration can be performed 
as
\begin{eqnarray}
\sigma_{xy}/H&\approx&
-\frac{e^3}{V_{0}}\sum_{\bf k}
\left(
-\frac{\partial f(E_{\bf k}^{-*})}{\partial E_{\bf k}^{-*}}
\right)
\left({a_{-,{\bf k}}^{\rm ff}}\right)^3
\frac{1}{2{\Gamma^{*}}^2}
\nonumber
\\
& &
\times
v_{{\bf k}x}
\left[
J_{{\bf k}x}\frac{\partial J_{{\bf k}y}}{\partial k_{y}}
-J_{{\bf k}y}\frac{\partial J_{{\bf k}x}}{\partial k_{y}}
\right],
\\
&=&
-\frac{e^3}{2V_{0}}\sum_{\bf k}
\delta\left(\mu-E_{\bf k}^{-*}
\right)
\frac{a_{-,{\bf k}}^{\rm ff}}{{\left[{\rm Im}\Sigma_{\bf k}^{\rm R}(\mu)\right]}^2}
\nonumber
\\
& &
\times
v_{{\bf k}x}
\left[
J_{{\bf k}x}\frac{\partial J_{{\bf k}y}}{\partial k_{y}}
-J_{{\bf k}y}\frac{\partial J_{{\bf k}x}}{\partial k_{y}}
\right], 
\label{eq:sxy_cancel}
\end{eqnarray}
where the last equation is derived for sufficiently low temperatures. 
Similarly to Eq.~(\ref{eq:sxx_cancel}), 
enhancement of the density of states of quasiparticles arises from $\delta(\mu-E_{\bf k}^{-*})$ 
by factor $z_{\bf k}^{-1}$, which cancels out by $a_{-,{\bf k}}^{\rm ff}$ 
 in Eq.~(\ref{eq:sxy_cancel}). 
Hence, the mass-enhancement factor does not appear in the expression of the Hall conductivity. 

As shown above, the formulas of $\sigma_{xx}$ and $\sigma_{xy}/H$ are obtained  
in the vicinity of the Fermi level located at the lower hybridized band 
on the basis of the Fermi liquid theory. 
In both expressions of $\sigma_{xx}$ and $\sigma_{xy}/H$, the renormalization factors of 
quasiparticles cancel out. 

In the next Sect., we will derive the exact formula of  
$\sigma_{xx}$ and $\sigma_{xy}/H$ for $U=0$ in Eq.~(\ref{eq:PAM}), 
which give the general formulas in the two-orbital systems, not restricted to 
the single band as treated in Eqs.~(\ref{eq:sxx_full}) and (\ref{eq:sxy_full}).
On the basis of the exactly-derived formulas, we will perform the explicit calculation 
of $\sigma_{xx}$ and $\sigma_{xy}/H$ in the periodic Anderson model 
with electron correlations to clarify the ground-state properties in Sect.4.

\section{Exact formulas of $\sigma_{xx}$ and $\sigma_{xy}$ for $U=0$}
\label{chap:U0}

In the case of $U=0$, Eq.~(\ref{eq:PAM}) is diagonalized as
\begin{eqnarray}
{\cal H}=\sum_{{\bf k}\sigma}
\left[
E_{\bf k}^{-}
\beta_{{\bf k}\sigma}^{\dagger}\beta_{{\bf k}\sigma}
+E_{\bf k}^{+}
\gamma_{{\bf k}\sigma}^{\dagger}\gamma_{{\bf k}\sigma}
\right],
\label{eq:PAM_diag}
\end{eqnarray}
where $E_{\bf k}^{-}$ is the lower hybridized band  
and $E_{\bf k}^{+}$ is the upper hybridized band, whose explicit form is given by 
\begin{eqnarray}
E_{\bf k}^{\mp}=\frac{\varepsilon_{\bf k}+\varepsilon_{\bf k}^{\rm f}}{2}
\mp\frac{\Delta_{\bf k}}{2}. 
\end{eqnarray}
Here, $\Delta_{\bf k}$ is defined as  
\begin{eqnarray}
\Delta_{\bf k}=\sqrt{\left(\varepsilon_{\bf k}-\varepsilon_{\bf k}^{\rm f}\right)^2+4V_{\bf k}^2}.   
\label{eq:Dlta}
\end{eqnarray}
Equation~(\ref{eq:PAM_diag}) is obtained by substituting 
$c_{{\bf k}\sigma}^{\dagger}=
u_{\bf k}\beta_{{\bf k}\sigma}^{\dagger}-w_{\bf k}\gamma_{{\bf k}\sigma}^{\dagger}$ 
and 
$f_{{\bf k}\sigma}^{\dagger}=
w_{\bf k}\beta_{{\bf k}\sigma}^{\dagger}+u_{\bf k}\gamma_{{\bf k}\sigma}^{\dagger}$ 
to Eq.~(\ref{eq:PAM}), 
where $u_{\bf k}$ and $w_{\bf k}$ satisfy
\begin{eqnarray}
u_{\bf k}^2&=&a_{\mp,{\bf k}}^{\rm cc}=\frac{1}{2}
\left(
1\mp\frac{\varepsilon_{\bf k}-\varepsilon_{\bf k}^{\rm f}}{\Delta_{\bf k}}
\right),
\label{eq:acc}
\\
w_{\bf k}^2&=&a_{\mp,{\bf k}}^{\rm ff}=\frac{1}{2}
\left(
1\pm\frac{\varepsilon_{\bf k}-\varepsilon_{\bf k}^{\rm f}}{\Delta_{\bf k}}
\right),  
\label{eq:aff}
\end{eqnarray}
respectively. 
Here, $a_{-,{\bf k}}^{\rm cc}$ $(a_{+,{\bf k}}^{\rm cc})$ represents 
the weight factor of the conduction electrons in the lower (upper) hybridized band 
and $a_{-,{\bf k}}^{\rm ff}$ $(a_{+,{\bf k}}^{\rm ff})$ represents 
the weight factor of the f electrons in the lower (upper) hybridized band. 
Note that in the case of $U=0$ in Eq.~(\ref{eq:PAM}), 
i.e., $\Sigma_{\bf k}^{\rm R}(\varepsilon)=0$,  
Eqs.~(\ref{eq:aff_U}) and (\ref{eq:acc_U}) reproduce Eqs.~(\ref{eq:aff}) and 
(\ref{eq:acc}), respectively. 

\subsection{Derivation of $\sigma_{xx}$ and $\sigma_{xy}$}

On the basis of the Kubo formula~\cite{Kubo}, 
the diagonal conductivity is given by
\begin{eqnarray}
\sigma_{xx}=\lim_{\omega\to 0}
\frac{\Phi_{xx}(\omega+i\delta)-\Phi_{xx}(0+i\delta)}{i\omega}.
\label{eq:sxx_kubo}
\end{eqnarray}
Here, the kernel $\Phi_{xx}$ is expressed as 
\begin{eqnarray}
\Phi_{xx}(i\omega_{m})=-e^2\frac{T}{V_0}\sum_{n}\sum_{{\bf k}\sigma}
{\rm Tr}
\left[
{\cal G}_{{\bf k}\sigma}(i\varepsilon_{n})
{\cal V}_{{\bf k}x}
{\cal G}_{{\bf k}\sigma}(i\varepsilon_{n}+i\omega_{m})
{\cal V}_{{\bf k}x}
\right]
\label{eq:Kernel_sxx}
\end{eqnarray}
with $\omega_{m}=2{\pi}mT$ and $\varepsilon_{n}=(2{\pi}+1)nT$ ($m$, $n$ are integers). 
Here, the velocity matrix and the Green-function matrix are given by 
\begin{eqnarray}
{\cal V}_{{\bf k}\eta}&=&
\left(
\begin{array}{cc}
v_{{\bf k}\eta}^{\rm c0}
&
\frac{\partial V_{\bf k}}{\partial k_{\eta}}
\\
\frac{\partial V_{\bf k}}{\partial k_{\eta}}
&
v_{{\bf k}\eta}^{\rm f0}
\end{array}
\right), 
\\
{\cal G}_{{\bf k}\sigma}(i\varepsilon_{n})
&=&
\left(
\begin{array}{cc}
G_{{\bf k}\sigma}^{\rm cc}(i\varepsilon_{n})
&
G_{{\bf k}\sigma}^{\rm cf}(i\varepsilon_{n})
\\
G_{{\bf k}\sigma}^{\rm fc}(i\varepsilon_{n})
&
G_{{\bf k}\sigma}^{\rm ff}(i\varepsilon_{n})
\end{array}
\right), 
\end{eqnarray}
respectively, where 
\begin{eqnarray}
G_{{\bf k}\sigma}^{\rm ff}(i\varepsilon_{n})&=&
\left[
i\varepsilon_{n}-\varepsilon_{\bf k}^{\rm f}-\frac{V_{\bf k}^
2}
{i\varepsilon_{n}-\varepsilon_{\bf k}}
\right]^{-1}, 
\\
G_{{\bf k}\sigma}^{\rm cc}(i\varepsilon_{n})&=&
\left[
i\varepsilon_{n}-\varepsilon_{\bf k}-\frac{V_{\bf k}^2}
{i\varepsilon_{n}-\varepsilon_{\bf k}^{\rm f}}
\right]^{-1}, 
\\
G_{{\bf k}\sigma}^{\rm cf}(i\varepsilon_{n})&=&
\frac{V_{\bf k}}{(i\varepsilon_{n}-\varepsilon_{\bf k}^{\rm f})
(i\varepsilon_{n}-\varepsilon_{\bf k})-V_{\bf k}^{2}}, 
\end{eqnarray}
and $G_{{\bf k}\sigma}^{\rm fc}(i\varepsilon_{n})
=G_{{\bf k}\sigma}^{\rm cf}(i\varepsilon_{n})$. 

By performing the analytic continuation, 
the conductivity in the periodic Anderson model for $U=0$ is derived as  
\begin{eqnarray}
\sigma_{xx}&=&\sigma_{xx}^{--}+\sigma_{xx}^{++}+\sigma_{xx}^{-+}+\sigma_{xx}^{+-},
\label{eq:sxx_tot}
\end{eqnarray}
where
\begin{eqnarray}
\sigma_{xx}^{--}&=&\frac{e^2}{V_{0}}\int_{-\infty}^{\infty}\frac{d\varepsilon}{\pi}
\left(
-\frac{\partial f(\varepsilon)}{\partial\varepsilon}
\right)
\sum_{{\bf k}\sigma}
\left(v_{{\bf k}x}^{--}\right)^2
\left\{
{\rm Im}G_{{\bf k}\sigma}^{- {\rm R}}(\varepsilon)
\right\}^2,
\label{eq:sxx_mm}
\\
\sigma_{xx}^{++}&=&\frac{e^2}{V_{0}}\int_{-\infty}^{\infty}\frac{d\varepsilon}{\pi}
\left(
-\frac{\partial f(\varepsilon)}{\partial\varepsilon}
\right)
\sum_{{\bf k}\sigma}
\left(v_{{\bf k}x}^{++}\right)^2
\left\{
{\rm Im}G_{{\bf k}\sigma}^{+ {\rm R}}(\varepsilon)
\right\}^2,
\label{eq:sxx_pp}
\\
\sigma_{xx}^{-+}+\sigma_{xx}^{+-}&=&
\frac{e^2}{V_{0}}\int_{-\infty}^{\infty}\frac{d\varepsilon}{\pi}
\left(
-\frac{\partial f(\varepsilon)}{\partial\varepsilon}
\right)
\sum_{{\bf k}\sigma}
v_{{\bf k}x}^{-+}v_{{\bf k}x}^{+-}
\nonumber
\\
& &\times
{\rm Re}\left[
G_{{\bf k}\sigma}^{- {\rm R}}(\varepsilon)G_{{\bf k}\sigma}^{+ {\rm A}}(\varepsilon)
-G_{{\bf k}\sigma}^{- {\rm R}}(\varepsilon)G_{{\bf k}\sigma}^{+ {\rm R}}(\varepsilon)
\right].  
\label{eq:sxx}
\end{eqnarray}
Here, $V_0$ is a volume of the system.
The velocity of the hybridized band 
$v_{{\bf k}\eta}^{\alpha\alpha}\equiv\partial E_{\bf k}^{(\alpha)}/\partial k_{\eta}$ 
$(\eta=x, y, z)$ is given by
\begin{eqnarray}
v_{{\bf k}\eta}^{\alpha\alpha}=
 a_{\alpha,{\bf k}}^{\rm cc}v_{{\bf k}\eta}^{\rm c0}
+a_{\alpha,{\bf k}}^{\rm ff}v_{{\bf k}\eta}^{\rm f0}
+a_{\alpha,{\bf k}}^{\rm cf}\frac{\partial V_{\bf k}}{\partial k_{\eta}}
+a_{\alpha,{\bf k}}^{\rm fc}\frac{\partial V_{\bf k}}{\partial k_{\eta}}, 
\label{eq:v_aa}
\end{eqnarray}
where 
$a_{\mp,{\bf k}}^{\rm cf}$ and $a_{\mp,{\bf k}}^{\rm fc}$ are defined as
\begin{eqnarray}
a_{-,{\bf k}}^{\rm cf}&=&-\frac{V_{\bf k}}{\Delta_{\bf k}}, \ \ \ \ 
a_{+,{\bf k}}^{\rm cf} = \frac{V_{\bf k}}{\Delta_{\bf k}},
\\
a_{-,{\bf k}}^{\rm fc}&=&-\frac{V_{\bf k}}{\Delta_{\bf k}}, \ \ \ \ 
a_{+,{\bf k}}^{\rm fc} = \frac{V_{\bf k}}{\Delta_{\bf k}},
\end{eqnarray}
respectively, which satisfies $a_{\mp,{\bf k}}^{\rm ff}+a_{\mp,{\bf k}}^{\rm cc}=1$ 
and $a_{\alpha,{\bf k}}^{\rm ff}a_{\alpha,{\bf k}}^{\rm cc}
=a_{\alpha,{\bf k}}^{\rm cf}a_{\alpha,{\bf k}}^{\rm fc}$. 
In Eq.~(\ref{eq:sxx}), the off-diagonal velocity is defined as
\begin{eqnarray}
v_{{\bf k}\eta}^{\alpha\overline{\alpha}}&=&
 \sqrt{a_{\alpha,{\bf k}}^{\rm cc}a_{\overline{\alpha},{\bf k}}^{\rm cc}}v_{{\bf k}\eta}^{\rm c0}
+\sqrt{a_{\alpha,{\bf k}}^{\rm ff}a_{\overline{\alpha},{\bf k}}^{\rm ff}}v_{{\bf k}\eta}^{\rm f0}
\nonumber
\\
&+&\sqrt{a_{\alpha,{\bf k}}^{\rm cc}a_{\overline{\alpha},{\bf k}}^{\rm ff}}\frac{\partial V_{\bf k}}{\partial k_{\eta}}
+\sqrt{a_{\alpha,{\bf k}}^{\rm ff}a_{\overline{\alpha},{\bf k}}^{\rm cc}}\frac{\partial V_{\bf k}}{\partial k_{\eta}}  
\label{eq:v_ab}
\end{eqnarray}
for $\overline{\alpha}=-\alpha$. 

The retarded Green function $G_{{\bf k}\sigma}^{\alpha {\rm R}}(\varepsilon)$
is given by
\begin{eqnarray}
G_{{\bf k}\sigma}^{\alpha {\rm R}}(\varepsilon)
=\frac{1}{\varepsilon-E^{(\alpha)}_{\bf k}+i\Gamma_{\bf k}^{(\alpha)}}.
\label{eq:GR}
\end{eqnarray}
The advanced Green function $G_{{\bf k}\sigma}^{\alpha{\rm A}}(\varepsilon)$ 
is obtained by the relation of 
$G_{{\bf k}\sigma}^{\alpha {\rm A}}(\varepsilon)=\left[
G_{{\bf k}\sigma}^{\alpha {\rm R}}(\varepsilon)
\right]^{*}$. 
Here, we introduce the imaginary part of the self energy $\Gamma_{\bf k}^{(\alpha)}$ 
in Eq.~(\ref{eq:GR}). 
This term can arise from the impurity scattering even in the $U=0$ periodic Anderson model. 
We discuss the general expression of $\sigma_{xx}$ and $\sigma_{xy}/H$ with 
the finite damping rate.

On the basis of the formalism in Refs.~\citen{FEW,Fukuyama}, 
the Hall conductivity in the periodic Anderson model for $U=0$ 
is derived. Hereafter, we show the result for $V_{\bf k}=V$ in Eq.~(\ref{eq:PAM}). 

\begin{figure}
\includegraphics[width=7.5cm]{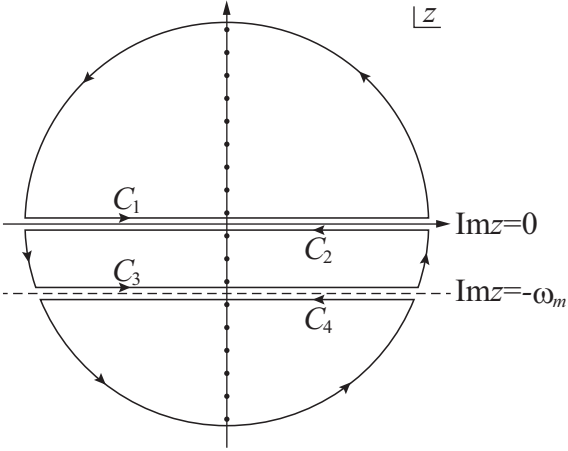}
\caption{
Contours of integration in the complex-$z$ plane. 
The dots on the imaginary axis represent the Fermionic thermal frequencies 
$i\varepsilon_{n}=i(2n+1)\pi T$ with $n$ being integers. 
}
\label{fig:z_plane}
\end{figure}

The Hall conductivity is given by
\begin{eqnarray}
\sigma_{xy}=\lim_{\omega\to 0}\lim_{{\bf q}\to{\bf 0}}
\frac{\Phi_{xy}({\bf q},\omega+i\delta)-\Phi_{xy}({\bf q},i\delta)}{i\omega}, 
\label{eq:sxy_kernel} 
\end{eqnarray}
where the kernel $\Phi_{xy}({\bf q},i\omega_m)$ is given by 
\begin{eqnarray}
\Phi_{xy}({\bf q},i\omega_m)=
(q_{x}A_{{\bf q}y}-q_{y}A_{{\bf q}x})e^3
\frac{T}{2V_0}\sum_{n}\sum_{{\bf k}\sigma}
{\rm Tr}
\left[
\right.
\nonumber
\\
\left.
{\cal V}_{{\bf k}x}
\left(
\frac{\partial}{\partial k_{x}}
{\cal G}_{{\bf k}\sigma}(i\varepsilon_{n})
\right)
\left(
\frac{\partial {\cal V}_{{\bf k}y}}{\partial k_{y}}
\right)
{\cal G}_{{\bf k}\sigma}(i\varepsilon_{n}+i\omega_{m})
\right.
\nonumber
\\
\left.
-{\cal V}_{{\bf k}x}
{\cal G}_{{\bf k}\sigma}(i\varepsilon_{n})
\left(\frac{\partial {\cal V}_{{\bf k}y}}{\partial k_{y}}\right)
\frac{\partial}{\partial k_{x}}
{\cal G}_{{\bf k}\sigma}(i\varepsilon_{n}+i\omega_{m})
\right.
\nonumber
\\
\left.
+
\left(v_{{\bf k}x}
\leftrightarrow v_{{\bf k}y}\right)
\right].  
\end{eqnarray}
Here $A_{{\bf q}\eta}$ is the $\eta$ component of the vector potential by which  
the magnetic field along the $z$ axis is expressed as 
$H=i\lim_{{\bf q}\to 0}(q_{x}A_{{\bf q}y}-q_{y}A_{{\bf q}x})$.
By performing the analytic continuation, 
the Hall conductivity is obtained as follows:  
\begin{eqnarray}
\sigma_{xy}=\sum_{\alpha=\mp}\sigma_{xy}^{\alpha\alpha}
+\sum_{\alpha=\mp}\sigma_{xy}^{\alpha\overline{\alpha}}
+\sigma_{xy}^{\rm extra},
\label{eq:sxy}
\end{eqnarray}
where 
\begin{eqnarray}
\sigma_{xy}^{\alpha\alpha}
&=&H\frac{e^3}{2V_{0}}\sum_{{\bf k}\sigma}
\int_{-\infty}^{\infty}\frac{d\varepsilon}{\pi}
\left(
-\frac{\partial f(\varepsilon)}{\partial\varepsilon}
\right)
\left|
G_{{\bf k}\sigma}^{\alpha {\rm R}}(\varepsilon)
\right|^2
{\rm Im}
\left[
G_{{\bf k}\sigma}^{\alpha {\rm R}}(\varepsilon)
\right]
\nonumber
\\
& &
\times
v_{{\bf k}x}^{\alpha\alpha}
\left(
 v_{{\bf k}x}^{\alpha\alpha}\frac{\partial v_{{\bf k}y}^{\alpha\alpha}}{\partial k_{y}}
-v_{{\bf k}y}^{\alpha\alpha}\frac{\partial v_{{\bf k}x}^{\alpha\alpha}}{\partial k_{y}}
\right),
\label{eq:sxy_1}
\\
%
\sigma_{xy}^{\alpha\overline{\alpha}}
&=&H\frac{e^3}{2V_{0}}\sum_{{\bf k}\sigma}
\int_{-\infty}^{\infty}\frac{d\varepsilon}{\pi}
\left(
-\frac{\partial f(\varepsilon)}{\partial\varepsilon}
\right)
{\rm Im}
\left[
\left\{
G_{{\bf k}\sigma}^{\alpha {\rm R}}(\varepsilon)
\right\}^2
G_{{\bf k}\sigma}^{\overline{\alpha} {\rm A}}(\varepsilon)
\right]
\nonumber
\\
& &
\times
v_{{\bf k}x}^{\alpha\alpha}
\left(
 v_{{\bf k}x}^{\alpha\overline{\alpha}}\frac{\partial v_{{\bf k}y}^{\overline{\alpha}\alpha}}{\partial k_{y}}
-v_{{\bf k}y}^{\alpha\overline{\alpha}}\frac{\partial v_{{\bf k}x}^{\overline{\alpha}\alpha}}{\partial k_{y}}
\right)
\nonumber
\\
&-&H\frac{e^3}{2V_{0}}\sum_{{\bf k}\sigma}
\int_{-\infty}^{\infty}\frac{d\varepsilon}{\pi}f(\varepsilon)
{\rm Im}
\left[
-2\left\{G_{{\bf k}\sigma}^{\alpha {\rm R}}(\varepsilon)\right\}^3
G_{{\bf k}\sigma}^{\overline{\alpha} {\rm R}}(\varepsilon)
\right.
\nonumber
\\
& &
\left.
+\left\{G_{{\bf k}\sigma}^{\alpha {\rm R}}(\varepsilon)\right\}^2
\left\{G_{{\bf k}\sigma}^{\overline{\alpha} {\rm R}}(\varepsilon)\right\}^2
\right]
v_{{\bf k}x}^{\alpha\alpha}
\left(
 v_{{\bf k}x}^{\alpha\overline{\alpha}}\frac{\partial v_{{\bf k}y}^{\overline{\alpha}\alpha}}{\partial k_{y}}
-v_{{\bf k}y}^{\alpha\overline{\alpha}}\frac{\partial v_{{\bf k}x}^{\overline{\alpha}\alpha}}{\partial k_{y}}
\right),
\label{eq:sxy_2}
\\
%
\sigma_{xy}^{\rm extra}
&=&H\frac{e^3}{2V_{0}}\sum_{{\bf k}\sigma}
\int_{-\infty}^{\infty}\frac{d\varepsilon}{\pi}
\left(
-\frac{\partial f(\varepsilon)}{\partial\varepsilon}
\right)
{\rm Im}
\left[
G_{{\bf k}\sigma}^{- {\rm R}}(\varepsilon)
G_{{\bf k}\sigma}^{+ {\rm A}}(\varepsilon)
\right]Q_{{\bf k}xy}
\nonumber
\\
&-&H\frac{e^3}{2V_{0}}\sum_{{\bf k}\sigma}
\int_{-\infty}^{\infty}\frac{d\varepsilon}{\pi}f(\varepsilon)
{\rm Im}
\left[
-\left\{G_{{\bf k}\sigma}^{- {\rm R}}(\varepsilon)\right\}^2
G_{{\bf k}\sigma}^{+ {\rm R}}(\varepsilon)
\right.
\nonumber
\\
& &
\left.
+G_{{\bf k}\sigma}^{- {\rm R}}(\varepsilon)
\left\{G_{{\bf k}\sigma}^{+{\rm R}}(\varepsilon)\right\}^2
\right]Q_{{\bf k}xy}.  
\label{eq:sxy3}
\end{eqnarray}
Here,  
$\overline{\alpha}=-\alpha$ and $Q_{{\bf k}xy}$ is defined by
\begin{eqnarray}
Q_{{\bf k}xy}&=&-2\frac{v_{{\bf k}x}^{\rm c0}-v_{{\bf k}x}^{\rm f0}}{\Delta_{\bf k}}
\frac{V^2}{\Delta_{\bf k}^2}
\nonumber
\\
&\times&
\left(
 v_{{\bf k}x}^{\rm c0}\frac{\partial v_{{\bf k}y}^{\rm c0}}{\partial k_{y}}
-v_{{\bf k}y}^{\rm c0}\frac{\partial v_{{\bf k}x}^{\rm c0}}{\partial k_{y}}
+v_{{\bf k}x}^{\rm f0}\frac{\partial v_{{\bf k}y}^{\rm f0}}{\partial k_{y}}
-v_{{\bf k}y}^{\rm f0}\frac{\partial v_{{\bf k}x}^{\rm f0}}{\partial k_{y}}
\right).
\label{eq:extra_factor}
\end{eqnarray}
In deriving Eq.~(\ref{eq:sxx_tot}) and Eq.~(\ref{eq:sxy}), the analytic continuation is performed as shown in Fig.~\ref{fig:z_plane}~\cite{Eliashberg,FEW,Kohno,Kontani2008,Narikiyo_I,Narikiyo_II,Fuseya}. Here, $C_1$ $(C_2)$ is the contour from $-\infty$ to $\infty$ (from $\infty$ to $-\infty$) at just above (below) ${\rm Im}z=0$ and $C_3$ $(C_4)$ is the contour from $-\infty$ to $\infty$ (from $\infty$ to $-\infty$) at just above (below) ${\rm Im}z=-\omega_m$. 
In Eq.~(\ref {eq:sxx_tot}) and Eq.~(\ref{eq:sxy}), the term including $f'(\varepsilon)$ term arises from the $C_2+C_3$ contours, which makes the main contribution come from the vicinity of the Femi level.
On the other hand, in Eq.~(\ref{eq:sxy}), 
the term including $f(\varepsilon)$ term arises from the contribution from the $C_1+C_4$ contours. 

It is noted that even when the hybridization has the momentum dependence, $V_{\bf k}$, 
Eqs.~(\ref{eq:sxy_1}) and (\ref{eq:sxy_2}) hold with the velocities defined 
in Eqs.~(\ref{eq:v_aa}) and (\ref{eq:v_ab}).
Equaions~(\ref{eq:sxx_tot}) and Eq.~(\ref{eq:sxy}) are the general expressions for 
$\sigma_{xx}$ and $\sigma_{xy}/H$, respectively, 
in the systems with the single-band as well as the two-bands at the Fermi level, 
which are constituted of two orbitals. 

To clarify the fundamental properties of the conductivity and Hall coefficient 
in the periodic Anderson model, hereafter we discuss the case of 
the flat dispersion of the f band, $\varepsilon_{\bf k}^{\rm f}=\varepsilon_{\rm f}$ 
in Eq.~(\ref{eq:PAM})
as the simplest typical case of the heavy-electron systems. 

An important remark is that in Eqs.~(\ref{eq:sxx_tot}) and (\ref{eq:sxy}) 
the velocities of the hybridized bands $v_{{\bf k}\eta}^{\alpha\alpha}$ appear, 
which give rise to the velocity of the ``large" Fermi surface 
which contains contributions from both f and conduction electrons, 
but not of the ``small"  Fermi surface for the conduction band.

\subsection{The limit of small damping rate at low temperatures}
\label{sec:small_Gamma}

When the total filling is less than the half-filling, i.e., $\bar{n}<2$ 
and the Fermi level is located at the lower hybridized band, 
$\sigma_{xx}=\sigma_{xx}^{--}$ holds in Eq.~(\ref{eq:sxx_tot}) 
at low temperatures 
for the small $\Gamma_{\bf k}^{-}$ to satisfy $\mu\tau_{\bf k}^{-}\gg 1$,  
where the relaxation time 
$\tau_{\bf k}^{-}$ is defined as $\tau_{\bf k}^{-}\equiv\frac{1}{2\Gamma_{\bf k}^{-}}$. 

In this case, from Eq.~(\ref{eq:sxx_mm}), we have  
\begin{eqnarray}
\sigma_{xx}\approx
\sigma_{xx}^{--}&=&\frac{e^2}{V_{0}}\sum_{{\bf k}\sigma}
\left(
-\frac{\partial f(E_{\bf k}^{-})}{\partial E_{\bf k}^{-}}
\right)
\left(v_{{\bf k}x}^{\rm c0}\right)^2
\left({a_{-,{\bf k}}^{\rm cc}}\right)^2
\nonumber
\\
& &
\times
\int_{-\infty}^{\infty}\frac{d\varepsilon}{\pi}
\left[
\frac{\Gamma_{\bf k}^{-}}{(\varepsilon-E_{\bf k}^{-})^2+{\Gamma_{\bf k}^{-}}^2}
\right]^2, 
\nonumber
\\
&=&
\frac{e^2}{V_{0}}\sum_{{\bf k}\sigma}
\left(
-\frac{\partial f(E_{\bf k}^{-})}{\partial E_{\bf k}^{-}}
\right)
\left(v_{{\bf k}x}^{\rm c0}\right)^2
\left({a_{-,{\bf k}}^{\rm cc}}\right)^2
\frac{1}{2\Gamma_{\bf k}^{-}}, 
\nonumber
\\
&=&
\frac{e^2}{V_{0}}\sum_{{\bf k}\sigma}
\left(
-\frac{\partial f(E_{\bf k}^{-})}{\partial E_{\bf k}^{-}}
\right)
\left(v_{{\bf k}x}^{\rm c0}\right)^2
\left({a_{-,{\bf k}}^{\rm cc}}\right)^2
\tau_{\bf k}^{-}.
\label{eq:sxx_Boltzmann}
\end{eqnarray}
Note that Eq.~(\ref{eq:sxx_full}) in the limit of $U=0$ reproduces the expression of 
Eq.~(\ref{eq:sxx_Boltzmann}). 
Since $-\frac{\partial f(E_{\bf k}^{-})}{\partial E_{\bf k}^{-}}=\delta(\mu-E_{\bf k}^{-})$ 
holds at $T=0$, we have
\begin{eqnarray}
\sigma_{xx}=\frac{e^2}{V_{0}}\sum_{{\bf k}\sigma}
\delta(\mu-E_{\bf k}^{-})
\left(v_{{\bf k}x}^{\rm c0}\right)^2
\left({a_{-,{\bf k}}^{\rm cc}}\right)^2
\tau_{\bf k}^{-}, 
\label{eq:sxx_T0}
\end{eqnarray}
in the ground state.

As for the Hall conductivity, in the small $\Gamma_{\bf k}^{-}$ limit at low temperatures, 
the contribution from the lower hybridized band $\sigma_{xy}^{--}$ in Eq.~(\ref{eq:sxy_1}) 
is only the relevant term for $\sigma_{xy}$. 
Then we have   
\begin{eqnarray}
\sigma_{xy}&\approx& H\frac{e^3}{2V_{0}}\sum_{{\bf k}\sigma}
\left(
-\frac{\partial f(E_{\bf k}^{-})}{\partial E_{\bf k}^{-}}
\right)
\left(v_{{\bf k}x}^{\rm c0}\right)^2
\left(
\frac{\partial v_{{\bf k}y}^{\rm c0}}{\partial k_{y}}
\right)
\left({a_{-,{\bf k}}^{\rm cc}}\right)^3
\nonumber
\\
& &
\times
\int_{-\infty}^{\infty}\frac{d\varepsilon}{\pi}
\left|
G_{{\bf k}\sigma}^{- {\rm R}}(\varepsilon)
\right|^2
{\rm Im}G_{{\bf k}\sigma}^{- {\rm R}}(\varepsilon), 
\nonumber
\\
&=&
-H\frac{e^3}{2V_{0}}\sum_{{\bf k}\sigma}
\left(
-\frac{\partial f(E_{\bf k}^{-})}{\partial E_{\bf k}^{-}}
\right)
\left(v_{{\bf k}x}^{\rm c0}\right)^2
\left(
\frac{\partial v_{{\bf k}y}^{\rm c0}}{\partial k_{y}}
\right)
\left({a_{-,{\bf k}}^{\rm cc}}\right)^3
2{\tau_{\bf k}^{-}}^2. 
\label{eq:sxy_simple}
\end{eqnarray}
Note that Eq.~(\ref{eq:sxy_full}) in the limit of $U=0$ reproduces the expression of Eq.~(\ref{eq:sxy_simple}).  
Since $-\frac{\partial f(E_{\bf k}^{-})}{\partial E_{\bf k}^{-}}=\delta(\mu-E_{\bf k}^{-})$ 
holds at $T=0$, we have
\begin{eqnarray}
\sigma_{xy}=-H\frac{e^3}{2V_{0}}\sum_{{\bf k}\sigma}
\delta(\mu-E_{\bf k}^{-})
\left(v_{{\bf k}x}^{\rm c0}\right)^2
\left(
\frac{\partial v_{{\bf k}y}^{\rm c0}}{\partial k_{y}}
\right)
\left({a_{-,{\bf k}}^{\rm cc}}\right)^3
2{\tau_{\bf k}^{-}}^2, 
\label{eq:sxy_T0}
\end{eqnarray}
in the ground state. 

\subsection{Isotropic free-electron system}

To analyze $\sigma_{xx}$ and $\sigma_{xy}$ explicitly, we consider the isotropic free electron system. 
Namely, the conduction electrons have the free dispersion as $\varepsilon_{\bf k}=\frac{k^2}{2m_c}$ 
in the periodic Anderson model in three spatial dimension. 
Here we assume $\tau=\tau_{\bf k}^{-}$ $(\Gamma=\Gamma_{\bf k}^{-})$ for simplicity of analysis.   
Equation~(\ref{eq:sxx_T0}) can be calculated as
\begin{eqnarray}
\sigma_{xx}=\frac{2e^2\tau}{(2\pi)^3}\int_{S(\mu)}dS
\frac{
\left(v_{{\bf k}x}^{\rm c0}\right)^2\left({a_{-,{\bf k}}^{\rm cc}}\right)^2
}{\left|\nabla E_{\bf k}^{-}\right|},
\label{eq:sxx_iso_1}
\end{eqnarray}
where the integral is taken as the surface integral over the constant-energy-surface $S(\mu)$ 
where $\mu=E_{\bf k}^{-}$ is satisfied in the $\bf k$ space~\cite{AM}. 
Since we have $\left|\nabla E_{\bf k}^{-}\right|={a_{-,{\bf k}}^{\rm cc}}v_{k}^{\rm c0}$ with 
$v_{k}^{\rm c0}\equiv\sqrt{(v_{{\bf k}x}^{\rm c0})^2+(v_{{\bf k}y}^{\rm c0})^2+(v_{{\bf k}z}^{\rm c0})^2}$ 
and $\int_{S(\mu)}dS=4\pi k_{\rm F}^2$ with 
$k_{\rm F}$ being the Fermi wave number in the isotropic free-electron system, 
Eq.~(\ref{eq:sxx_iso_1}) leads to 
\begin{eqnarray}
\sigma_{xx}&=&\frac{2e^2\tau}{(2\pi)^3}4\pi k_{\rm F}^2\
\frac{
\left(v_{{k_{\rm F}}x}^{\rm c0}\right)^2{a_{-,{k_{\rm F}}}^{\rm cc}}
}{v_{k_{{\rm F}}}^{\rm c0}}, 
\\
&=&\frac{\bar{n}e^2\tau}{m_{\rm c}}{a_{-,k_{\rm F}}^{\rm cc}},  
\label{eq:sxx_T0_iso} 
\end{eqnarray}
where $v_{k_{\rm F}}^{\rm c0}=k_{\rm F}/m_{\rm c}$, 
$v_{k_{\rm F}x}^{\rm c0}=k_{{\rm F}x}/m_{\rm c}$, 
and $k_{{\rm F}x}^2=k_{\rm F}^2/3$ are used. 
Here, $\bar{n}$ is the total filling, which is defined by 
the total electron number $N_{\rm e}$ per the volume of the system: 
\begin{eqnarray}
\bar{n}=\frac{N_{\rm e}}{V_{0}}
=2\frac{4\pi}{(2\pi)^3}\int_{0}^{k_{\rm F}}dkk^2
=\frac{k_{\rm F}^3}{3\pi^2}.
\end{eqnarray}
Note that the factor $\tau a_{-,k_{\rm F}}^{\rm cc}$ appears in Eq.~(\ref{eq:sxx_T0_iso}), 
which implies that the ratio of the amplitude of conduction electrons to the damping rate 
determines the nature of $\sigma_{xx}$.

As for the Hall conductivity, Eq.~(\ref{eq:sxy_T0}) can be calculated as
\begin{eqnarray}
\sigma_{xy}&=&-H\frac{2e^3\tau^2}{(2\pi)^3m_{\rm c}}\int_{S(\mu)}dS
\frac{
\left(v_{{\bf k}x}^{\rm c0}\right)^2
\left({a_{-,{\bf k}}^{\rm cc}}\right)^3
}
{\left|\nabla E_{\bf k}^{-}\right|}, 
\label{eq:sxy_ds}
\\
&=&
-H\frac{2e^3\tau^2}{(2\pi)^3m_{\rm c}}4\pi k_{\rm F}^2
\frac{
\left(v_{k_{{\rm F}x}}^{\rm c0}\right)^2
\left({a_{-,{k_{\rm F}}}^{\rm cc}}\right)^2
}
{v_{k_{\rm F}}},
\label{eq:sxy_T0_iso_2}
\\
&=&
-\omega_{\rm c}\tau\sigma_{xx}{a_{-,{k_{\rm F}}}^{\rm cc}}, 
\label{eq:sxy_T0_iso}
\end{eqnarray}
where $\omega_{c}$ is the cyclotron frequency defined as 
$\omega_{\rm c}\equiv\frac{eH}{m_{\rm c}}$. 
Note that in Eq.~(\ref{eq:sxy_T0_iso_2}) the factor     
$\left(\tau a_{-,k_{\rm F}}^{\rm cc}\right)^2$
appears, which implies that the square of the ratio of the amplitude of conduction electrons to the damping rate determines the nature of $\sigma_{xy}$.

By using Eqs.~(\ref{eq:sxx_T0_iso}) and (\ref{eq:sxy_T0_iso}), 
the Hall coefficient $R_{\rm H}$ 
under a weak magnetic field $H$ applied along the $z$ axis 
is obtained as
\begin{eqnarray}
R_{\rm H}&=&\frac{\sigma_{xy}/H}{\sigma_{xx}^2}, 
\\
&=&
-\frac{\omega_{\rm c}\tau a_{-,k_{\rm F}}^{\rm cc}}{H\sigma_{xx}}, 
\\
&=&
-\frac{1}{\bar{n}e}.  
\label{eq:1ne}
\end{eqnarray}
Note here that although both $\sigma_{xx}$ and $\sigma_{xy}$ include the ratio of 
the weight factor 
$a_{-,k_{\rm F}}^{\rm cc}$ 
to the damping rate as Eqs.~(\ref{eq:sxx_T0_iso}) and (\ref{eq:sxy_T0_iso}), 
the factors $\tau a_{-,k_{\rm F}}^{\rm cc}$ in the Hall coefficient are cancelled 
so that the resultant $R_{\rm H}$ is expressed by the total electron filling $\bar{n}$. 
This implies that $R_{\rm H}$ is only determined by the total filling 
irrespective of the weight of conduction electrons component at the Fermi level. 
Namely, Eq.~(\ref{eq:1ne}) reproduces the well-known result in the single-orbital system.

Here, two remarks should be made. 
First, Eq.~(\ref{eq:1ne}) shows the negative sign and 
that the magnitude is expressed as inverse of the total filling. 
We note that Eq.~(\ref{eq:1ne}) 
is obtained in the free-electron system with the spherical Fermi surface. 
In general, 
$R_{\rm H}$ depends on the shape of the Fermi surface, more precisely, the curvature of the 
Fermi surface. 
Hence, sign of $R_{\rm H}$ and the magnitude itself depend on the shape of the Fermi surface 
even in the small-$\Gamma_{\bf k}^{-}$ limit at $T=0$. 
This point will be discussed in detail in Sect.~\ref{sec:sxx_RH}. 

Second, we note that 
Eq.~(\ref{eq:1ne}) is obtained in the small-$\Gamma_{\bf k}^{-}$ limit. 
If the damping rate $\Gamma_{\bf k}^{-}$ is not small, 
it is not guaranteed that 
$R_{\rm H}$ shows a constant behavior as Eq.~(\ref{eq:1ne}) 
when $\varepsilon_{\rm f}$ is varied even at $T=0$, as will be discussed  
in Sect.~\ref{sec:Gm_dep}.

\section{Ground-state properties of $\sigma_{xx}$, $\sigma_{xy}/H$, and $R_{\rm H}$ 
on the square lattice}
\label{chap:SL}

In Sect.~3, we derived exactly the general expressions of diagonal and Hall conductivities in hybridized two-orbital systems with arbitrary band dispersions for non-interacting case. In this Sect. we study the ground-state properties of the diagonal conductivity and normal Hall effect in the periodic Anderson model with onsite Coulomb repulsion between f electrons. By employing the Fermi liquid theory discussed in Sect.~2, we will discuss that the diagonal and Hall conductivities can be calculated by using the formulas derived in Sect.~3. 
However, we make an approximation in which the Fermi liquid correction for the current given by Eq.~(\ref{eq:J}) is neglected. 
Nevertheless, this approximation is considered to be valid for the present purpose that we discuss a qualitative aspect of the diagonal and Hall conductivities. 
To clarify the general properties realized in Ce- and Yb-based heavy-electron systems, 
we concentrate on the Fermi-liquid ground state 
taking into account the effect of weak impurity scatterings. 

The imaginary part of the f-electron self energy around the Fermi level is expressed as
\begin{eqnarray}
{\rm Im}\Sigma_{\bf k}^{\rm R}(\varepsilon)
={\rm Im}\Sigma^{U \ {\rm R}}_{\bf k}(\varepsilon)+{\rm Im}\Sigma^{\rm imp \ R}, 
\label{eq:ImS_general}
\end{eqnarray}
where ${\rm Im}\Sigma^{U \ {\rm R}}_{\bf k}(\varepsilon)$ is arising from the onsite Coulomb repulsion 
in Eq.~(\ref{eq:PAM}) and ${\rm Im}\Sigma^{\rm imp \ {\rm R}}$ is from the impurity scattering. 
In the Fermi-liquid regime, ${\rm Im}\Sigma^{U \ {\rm R}}_{\bf k}(\varepsilon)$ has 
the following form at zero temperature: 
\begin{eqnarray}
{\rm Im}\Sigma^{U \ {\rm R}}_{\bf k}(\varepsilon)=-C_{U}(\varepsilon-\mu)^2, 
\end{eqnarray}
where $C_{U}>0$ is a constant of the order of the inverse of 
the effective Fermi energy~\cite{AGD,Yamada}.
When f electrons are scattered weakly by a small amount of local impurities,  
${\rm Im}\Sigma^{\rm imp \ R}$ is calculated within the Born approximation~\cite{AGD,
MK2005
} as 
\begin{eqnarray}
{\rm Im}\Sigma^{\rm imp \ R}\approx-\pi n_{\rm imp}u^2\langle a^{\rm ff}_{-,{{\bf k}_{\rm F}}}\rangle_{\rm av}N^{*}(\mu),
\label{eq:S_imp}
\end{eqnarray}
where $n_{\rm imp}$ is the impurity concentration and $u$ is the impurity potential. 
Here, 
$N^{*}(\mu)$ is the density of states of the quasiparticles at the Fermi level defined by $N^{*}(\varepsilon)\equiv\sum_{\bf k}\delta(\varepsilon-E^{-*}_{\bf k})/N$  with $N$ being the number of lattice sites and $\langle a^{\rm ff}_{-,{{\bf k}_{\rm F}}}\rangle_{\rm av}$ is averaged value of the f-electron weight factor defined in Eq.~(\ref{eq:aff_U}) over the Fermi level. 
Although $N^{*}(\mu)$ is enhanced by the renormalization factor $z_{{\bf k}_{\rm F}}^{-1}$, the enhancement is canceled by the factor $\langle a^{\rm ff}_{-,{\bf k}}\rangle_{\rm av}$ [see Eq.~(\ref{eq:aff_U})]. Hence, $\langle a^{\rm ff}_{-,{{\bf k}_{\rm F}}}\rangle_{\rm av}N^{*}(\mu)$ is the quantity in the order of $(\pi V^2N_{\rm cF})^{-1}$, where $N_{\rm cF}$ is the density of states of conduction electrons at the Fermi level.  

As described in the formalism in Sect.~\ref{sec:PAM_finiteU}, 
the diagonal conductivity in Eq.~(\ref{eq:sxx_full}) and Hall conductivity 
in Eq.~(\ref{eq:sxy_full}) are claculated by using the Green function for quasiparticles. 
Here we consider the ${\rm Im}\Sigma_{\bf k}^{\rm R}(\mu)$  
in the form of Eq.~(\ref{eq:ImS_general}) as the self energy.
Since the impurity concentration $n_{\rm imp}$ and the strength of the impurity potential 
$u$ are parameters to be given 
and the extra factor $\langle a^{\rm ff}_{-,{{\bf k}_{\rm F}}}\rangle_{\rm av}N^{*}(\mu)$ in Eq.~(\ref{eq:S_imp}) can be expressed essentially by bare quantities not including renormalization factor 
and has only weak dependence in $\varepsilon_{\rm f}$
,
 we treat them as variable input parameters. 
Namely, we calculate $\sigma_{xx}$ and $\sigma_{xy}/H$ by inputting 
\begin{eqnarray}
\Gamma\equiv-{\rm Im}\Sigma_{\bf k}^{\rm R}(\mu)=-{\rm Im}\Sigma^{\rm imp \ R} 
\label{eq:Gamma_def}
\end{eqnarray}
into the quasiparticle Green function. 
As the simplest framework to perform such a calculation, we employ the slave-boson 
mean field theory~\cite{KR} since it has been established to describe the fixed point of 
the Fermi-liquid ground state in the periodic Anderson model correctly.

\subsection{Slave-boson mean field theory}
\label{sec:sbMF}

By applying the slave-boson mean field theory~\cite{KR} to Eq.~(\ref{eq:PAM}), 
the effective Hamiltonian is obtained as  
\begin{eqnarray}
\tilde{\cal H}&=&\sum_{{\bf k}\sigma}\varepsilon_{\bf k}c_{{\bf k}\sigma}^{\dagger}c_{{\bf k}\sigma}
+\sum_{i\sigma}\varepsilon_{\rm f}
f_{i\sigma}^{\dagger}f_{i\sigma}
\nonumber
\\
&+&V\sum_{i\sigma}\left(z_{i\sigma}f_{i\sigma}^{\dagger}c_{i\sigma}+
c_{i\sigma}^{\dagger}f_{i\sigma}z_{i\sigma}^{\dagger}
\right)
+U\sum_{i}d_{i}^{\dagger}d_{i}
\nonumber
\\
&+&\sum_{i}\lambda_{i}^{(1)}\left(
e_{i}^{\dagger}e_{i}+p_{i\sigma}^{\uparrow}p_{i\uparrow}+p_{i\downarrow}^{\dagger}p_{i\downarrow}
+d_{i}^{\dagger}d_{i}-1
\right)
\nonumber
\\
&+&\sum_{i\sigma}\lambda_{i\sigma}^{(2)}\left(
f_{i\sigma}^{\dagger}f_{i\sigma}-p_{i\sigma}^{\dagger}p_{i\sigma}-d_{i}^{\dagger}d_{i}
\right). 
\label{eq:PAM_MF}
\end{eqnarray}
Here, $e_{i}^{\dagger}$ $(e_{i})$ and $d_{i}^{\dagger}$ $(d_{i})$ are bose creation (annihilation) 
operators for the empty and doubly-occupied state, respectively 
and $p_{i\sigma}^{\dagger}$ $(p_{i\sigma})$ for the singly-occupied state on the $i$-th site. 
The renormalization factor is defined as
\begin{eqnarray}
z_{i\sigma}&\equiv&\left(1-d_{i}^{\dagger}d_{i}-p_{i\sigma}^{\dagger}p_{i\sigma}\right)^{-1/2}
\left(e_{i}^{\dagger}p_{i\sigma}+p_{i-\sigma}^{\dagger}d_{i}\right)
\nonumber
\\
& &
\times
\left(1-e_{i}^{\dagger}e_{i}-p_{i-\sigma}^{\dagger}p_{i-\sigma}\right)^{-1/2}.
\label{eq:z_MF}
\end{eqnarray}
The last two terms with the Lagrange multipliers $\lambda_{i}^{(1)}$ and $\lambda_{i}^{(2)}$ 
in Eq.~(\ref{eq:PAM_MF}) 
are introduced to require the constraint for the completeness condition.

Here, we consider the case where f electrons are subject to the impurity scattering in the form of Eq.~(\ref {eq:Gamma_def}) as an input parameter.  

By approximating the mean fields as uniform ones,  
$e=\langle e_{i}\rangle$, $p_{\sigma}=\langle p_{i\sigma}\rangle$, $d=\langle d_{i}\rangle$, 
and the Lagrange multipliers,  
$\lambda^{(1)}=\lambda_{i}^{(1)}$ and $\lambda^{(2)}=\lambda_{i}^{(2)}$, 
with $z_{\sigma}=\langle z_{i\sigma}\rangle$, 
the set of mean-field equations is obtained by 
$\partial\langle \tilde{\cal H}\rangle/\partial\lambda^{(1)}=0$, $\partial\langle \tilde{\cal H}\rangle/\partial\lambda^{(2)}=0$, 
$\partial\langle \tilde{\cal H}\rangle/\partial p_{\uparrow}=0$, $\partial\langle \tilde{\cal H}\rangle/\partial p_{\downarrow}=0$,  
$\partial\langle \tilde{\cal H}\rangle/\partial d=0$, and
$\partial\langle \tilde{\cal H}\rangle/\partial e=0$: 
\begin{eqnarray}
e^2+p_{\uparrow}^2+p_{\downarrow}^2+d^2&=&1, 
\label{eq:MF1}
\\
\frac{1}{N}\sum_{{\bf k}\sigma}\langle f_{{\bf k}\sigma}^{\dagger}f_{{\bf k}\sigma}\rangle
&=&p_{\uparrow}^2+p_{\downarrow}^2, 
\label{eq:MF2}
\\
\frac{V}{N}\sum_{{\bf k}\sigma}
\left(\frac{\partial z_{\sigma}}{\partial p_{\uparrow}}\right)
\left(
\langle f_{{\bf k}\sigma}^{\dagger}c_{{\bf k}\sigma}\rangle
+{\rm h.c.}
\right)
& &
\nonumber
\\
+2\sum_{{\bf k}\sigma}z_{\sigma}\left(\frac{\partial z_{\sigma}}{\partial p_{\uparrow}}\right)
\varepsilon_{\rm f}
\langle
f^{\dagger}_{{\bf k}\sigma}f_{{\bf k}\sigma}
\rangle
&=&-2\left(\lambda^{(1)}-\lambda^{(2)} \right)p_{\uparrow},
\label{eq:MF3}
\\
\frac{V}{N}\sum_{{\bf k}\sigma}
\left(\frac{\partial z_{\sigma}}{\partial p_{\downarrow}}\right)
\left(
\langle f_{{\bf k}\sigma}^{\dagger}c_{{\bf k}\sigma}\rangle
+{\rm h.c.}
\right)
& &
\nonumber
\\
+2\sum_{{\bf k}\sigma}z_{\sigma}\left(\frac{\partial z_{\sigma}}{\partial p_{\downarrow}}\right)
\varepsilon_{\rm f}
\langle
f^{\dagger}_{{\bf k}\sigma}f_{{\bf k}\sigma}
\rangle
&=&-2\left(\lambda^{(1)}-\lambda^{(2)} \right)p_{\downarrow}, 
\label{eq:MF4}
\\
\frac{V}{N}\sum_{{\bf k}\sigma}
\left(\frac{\partial z_{\sigma}}{\partial d}\right)
\left(
\langle f_{{\bf k}\sigma}^{\dagger}c_{{\bf k}\sigma}\rangle
+{\rm h.c.}
\right)
& &
\nonumber
\\
+2\sum_{{\bf k}\sigma}z_{\sigma}\left(\frac{\partial z_{\sigma}}{\partial d}\right)
\varepsilon_{\rm f}
\langle
f^{\dagger}_{{\bf k}\sigma}f_{{\bf k}\sigma}
\rangle
&=&-2\left(\lambda^{(1)}-2\lambda^{(2)} \right)d, 
\label{eq:MF5}
\\
\frac{V}{N}\sum_{{\bf k}\sigma}
\left(\frac{\partial z_{\sigma}}{\partial e}\right)
\left(
\langle f_{{\bf k}\sigma}^{\dagger}c_{{\bf k}\sigma}\rangle
+{\rm h.c.}
\right)
& &
\nonumber
\\
+2\sum_{{\bf k}\sigma}z_{\sigma}\left(\frac{\partial z_{\sigma}}{\partial e}\right)
\varepsilon_{\rm f}
\langle
f^{\dagger}_{{\bf k}\sigma}f_{{\bf k}\sigma}
\rangle
&=&-2\lambda^{(1)}e.  
\label{eq:MF6}
\end{eqnarray}
In this paper we consider the paramagnetic state and hence we assume 
$p_{\uparrow}=p_{\downarrow}=p$.  
The renormalization factor is then expressed as
\begin{eqnarray}
z_{\uparrow}=z_{\downarrow}=z=\frac{\sqrt{1-2p^2}}{1-p^2}.
\end{eqnarray}
Hereafter, we consider the case for $U=\infty$ for simplicity of analysis. 
Then we set $d=0$. 
By using Eq.~(\ref{eq:MF1}), 
the derivative of the renormalization factor by $p_{\sigma}$ and $e$ 
can be expressed by only $p$, as follows: 
\begin{eqnarray} 
\frac{\partial z_{\sigma}}{\partial p_{\sigma}}&=&\frac{\sqrt{1-2p^2}}{p\left(1-p^2\right)^{3/2}},
\\
\frac{\partial z_{\sigma}}{\partial p_{-\sigma}}&=&\frac{\sqrt{1-2p^2}}{p\sqrt{1-p^2}},
\\
\frac{\partial z_{\sigma}}{\partial e}&=&\frac{\sqrt{1-p^2}}{p^2}. 
\end{eqnarray}
Then, we end up by solving  
the mean-field equations of Eqs.~(\ref{eq:MF2}), (\ref{eq:MF3}), and (\ref{eq:MF5}) 
for $\lambda^{(1)}$, $\lambda^{(2)}$, and $p$ simultaneously with 
Eq.~(\ref{eq:filling_PAM}) for the chemical potential $\mu$ 
in the self-consistent manner. 
In these equations, the following expectation values are calculated as 
\begin{eqnarray}
\frac{1}{N}\sum_{{\bf k}\sigma}\langle f_{{\bf k}\sigma}^{\dagger}f_{{\bf k}\sigma}\rangle
&=&-\frac{1}{\pi N}\sum_{{\bf k}\sigma}\int_{-\infty}^{\infty}d\varepsilon f(\varepsilon)
{\rm Im}G_{{\bf k}\sigma}^{\rm ff \ R}(\varepsilon),
\\
\frac{1}{N}\sum_{{\bf k}\sigma}\langle c_{{\bf k}\sigma}^{\dagger}c_{{\bf k}\sigma}\rangle
&=&-\frac{1}{\pi N}\sum_{{\bf k}\sigma}\int_{-\infty}^{\infty}d\varepsilon f(\varepsilon)
{\rm Im}G_{{\bf k}\sigma}^{\rm cc \ R}(\varepsilon),
\\
\frac{1}{N}\sum_{{\bf k}\sigma}\langle f_{{\bf k}\sigma}^{\dagger}c_{{\bf k}\sigma}\rangle
&=&-\frac{1}{\pi N}\sum_{{\bf k}\sigma}\int_{-\infty}^{\infty}d\varepsilon f(\varepsilon)
{\rm Im}G_{{\bf k}\sigma}^{\rm fc \ R}(\varepsilon).  
\end{eqnarray}
%
Here, $G_{{\bf k}\sigma}^{\rm ff \ R}(\varepsilon)$, 
$G_{{\bf k}\sigma}^{\rm cc \ R}(\varepsilon)$, and 
$G_{{\bf k}\sigma}^{\rm fc \ R}(\varepsilon)$ are the retarded 
f-electron, conduction-electron, and off-diagonal green functions, respectively, 
which are given by 
\begin{eqnarray}
G_{{\bf k}\sigma}^{\rm ff \ {\rm R}}(\varepsilon)&=&
\tilde{a}_{-,{\bf k}}^{\rm ff}
\tilde{G}_{{\bf k}\sigma}^{- {\rm R}}(\varepsilon)
+\tilde{a}_{+,{\bf k}}^{\rm ff}
\tilde{G}_{{\bf k}\sigma}^{+ {\rm R}}(\varepsilon),
\label{eq:Gff}
\\
G_{{\bf k}\sigma}^{\rm cc \ {\rm R}}(\varepsilon)&=&
\tilde{a}_{-,{\bf k}}^{\rm cc}
\tilde{G}_{{\bf k}\sigma}^{- {\rm R}}(\varepsilon)
+\tilde{a}_{+,{\bf k}}^{\rm cc}
\tilde{G}_{{\bf k}\sigma}^{+ {\rm R}}(\varepsilon),
\label{eq:Gcc}
\\
G_{{\bf k}\sigma}^{\rm fc \ {\rm R}}(\varepsilon)&=&
\tilde{a}_{-,{\bf k}}^{\rm fc}
\tilde{G}_{{\bf k}\sigma}^{- {\rm R}}(\varepsilon)
+\tilde{a}_{+,{\bf k}}^{\rm fc}
\tilde{G}_{{\bf k}\sigma}^{+ {\rm R}}(\varepsilon),  
\label{eq:Gfc}
\end{eqnarray}
respectively. 
In Eq.~(\ref{eq:Gff}), $\tilde{a}_{-,{\bf k}}^{\rm ff}$ $(\tilde{a}_{+,{\bf k}}^{\rm ff})$ is the amplitude of the f-electron component
in the lower (upper)-hybridized band at ${\bf k}$.  
In Eq.~(\ref{eq:Gcc}), $\tilde{a}_{-,{\bf k}}^{\rm cc}$ $(\tilde{a}_{+,{\bf k}}^{\rm cc})$ is the amplitude of the conduction-electron component
in the lower (upper)-hybridized band at ${\bf k}$. 
These are given by 
\begin{eqnarray}
\tilde{a}_{\pm,{\bf k}}^{\rm ff}=
\tilde{a}_{\mp,{\bf k}}^{\rm cc}&=&\frac{1}{2}
\left(
1\mp\frac{\varepsilon_{\bf k}-\tilde{\varepsilon}_{\rm f}}{\tilde{\Delta}_{\bf k}}
\right). 
\label{eq:acc_renm}
\end{eqnarray}
The weight factor $\tilde{a}_{-,{\bf k}}^{\rm fc}$ $(\tilde{a}_{+,{\bf k}}^{\rm fc})$ 
in the lower (upper) hybridized band in Eq.~(\ref{eq:Gfc}) is given by
\begin{eqnarray}
\tilde{a}_{-,{\bf k}}^{\rm fc}&=&-\frac{\tilde{V}}{\tilde{\Delta}_{\bf k}},
\\
\tilde{a}_{+,{\bf k}}^{\rm fc}&=&\frac{\tilde{V}}{\tilde{\Delta}_{\bf k}}. 
\end{eqnarray}
Here, $\tilde{\Delta}_{\bf k}$ is given by 
\begin{eqnarray}
\tilde{\Delta}_{\bf k}=\sqrt{(\varepsilon_{\bf k}-\tilde{\varepsilon}_{\rm f})^2+4\tilde{V}^2}, 
\label{eq:Dlt_renm}
\end{eqnarray}
where $\tilde{\varepsilon}_{\rm f}\equiv\varepsilon_{\rm f}+\lambda^{(2)}$ 
and $\tilde{V}\equiv Vz$. 
We note that the relation $\tilde{a}_{\alpha,{\bf k}}^{\rm ff}+\tilde{a}_{\alpha,{\bf k}}^{\rm cc}=1$ holds. 
In Eqs.~(\ref{eq:acc_renm}) and (\ref{eq:Dlt_renm}), 
the f level $\varepsilon_{\rm f}$ and the hybridization strength $V$ are replaced by 
$\tilde{\varepsilon}_{\rm f}$ and $\tilde{V}$ in Eqs.~(\ref{eq:acc}) 
and (\ref{eq:Dlta}), respectively. 
The retarded Green function $\tilde{G}_{{\bf k}\sigma}^{\alpha {\rm R}}(\varepsilon)$ is given by 
\begin{eqnarray}
\tilde{G}_{{\bf k}\sigma}^{\alpha {\rm R}}(\varepsilon)
=\frac{1}{\varepsilon-\tilde{E}^{(\alpha)}_{\bf k}+i\tilde{\Gamma}_{\bf k}^{(\alpha)}}, 
\label{eq:GR_renm}
\end{eqnarray}
where $\tilde{E}^{(\alpha)}_{\bf k}$ is given by
\begin{eqnarray}
\tilde{E}_{\bf k}^{\mp}=\frac{\varepsilon_{\bf k}+\tilde{\varepsilon}_{\rm f}}{2}
\mp\frac{\tilde{\Delta}_{\bf k}}{2}. 
\label{eq:Emband}
\end{eqnarray}
Here we consider the finite imaginary part of the self energy, $\tilde{\Gamma}_{\bf k}^{(\alpha)}$, 
in Eq.~(\ref{eq:GR_renm}), as described in Eq.~(\ref{eq:Gamma_def}). 
In Sect.~\ref{sec:PAM_finiteU},
Eq.~(\ref{eq:aff_U}) is expressed as 
\begin{eqnarray}
a_{\alpha,{\bf k}}^{\rm ff}=
z_{\bf k}\left.\left[1+\frac{z_{\bf k}V^2}
{(\varepsilon-\varepsilon_{\bf k})^2}\right]^{-1}\right|_{\varepsilon=E_{\bf k}^{\alpha*}}. 
\label{eq:aff_2}
\end{eqnarray}
In the present mean-field framework, 
the renormalization factor is expressed as ${z}$ and
the quasiparticle band is expressed as $\tilde{E}_{\bf k}^{(\alpha)}$. 
Hence, by setting 
$z_{\bf k}$ as ${z}$ and $E_{\bf k}^{\alpha*}$ as $\tilde{E}_{\bf k}^{(\alpha)}$ 
in Eq.~(\ref{eq:aff_2}),  
$a_{\alpha,{\bf k}}^{\rm ff}$ is expressed as ${z}\tilde{a}_{\alpha,{\bf k}}^{\rm ff}$. 
Then, from Eq.~(\ref{eq:Gamma_renm}), the damping rate of the quasiparticle is expressed as 
\begin{eqnarray}
\tilde{\Gamma}_{\bf k}^{(\alpha)}={z}\tilde{a}_{\alpha,{\bf k}}^{\rm ff}\Gamma, 
\label{eq:damping_MF}
\end{eqnarray}
where $\Gamma$ is defined as $\Gamma\equiv-{\rm Im}\Sigma_{\bf k}^{\rm R}(\mu)>0$ 
in Eq.~(\ref{eq:Gamma_def}). 

Hence, 
by using Eq.~(\ref{eq:GR_renm}) as the Green function for quasiparticles, 
the ground-state properties of $\sigma_{xx}$, $\sigma_{xy}/H$, 
and the Hall coefficient 
will be discussed in the next Sect. 
on the basis of the exactly-derived formulas, Eq.~(\ref{eq:sxx_tot}) and Eq.~(\ref{eq:sxy}).

As shown in Ref.~\citen{Yamada}, by calculating the vertex correction in the self energy and the total current  consistently for the conductivity in the clean limit at finite temperatures, the total current $J_{{\bf k}\eta}$ has a finite value without diverging because of the presence of the Umklapp process in the periodic lattice in Eq.~(\ref{eq:sxx_cancel}). 
In the present framework for the ground state $(T=0)$, 
we consider the self energy with the impurity scattering as in Eq.~(\ref{eq:ImS_general}), which is consequently expressed as Eq.~(\ref{eq:Gamma_def}). 
As for the total current, 
the present framework corresponds to approximating the resultant $J_{{\bf k }\eta}$ as $v_{{\bf k}\eta}$ in Eq.~(\ref{eq:J}). 

The validity of this framework at least within approximating $J_{{\bf k}\eta}$ as $v_{{\bf k}\eta}$ is confirmed by comparing it with the finite-$U$ result 
based on the Fermi-liquid theory in Sect.~\ref{sec:PAM_finiteU}, 
which will be shown below (see also Appendix).

\subsection{Numerical Results}

On the basis of the theoretical framework described in the previous Sect., 
we calculate the conductivity in the periodic Anderson model on the square lattice. 
We consider 
the nearest-neighbor hopping for conduction electrons on the square lattice and 
the energy band is  
given by $\varepsilon_{\bf k}=-2t[\cos(k_x)+\cos(k_y)]$. 
As a typical parameter for heavy electrons, we set $t=1$, $V=0.3$, $U=\infty$ 
at the filling $\bar{n}=7/4$.  Hereafter, the transfer of conduction electrons is taken 
to be the energy unit of the parameters in the Hamiltonian, Eq.~(\ref{eq:PAM_MF}). 
The imaginary part of the f-electron self energy is set to be $\Gamma=10^{-3}$ 
in Eq.~(\ref{eq:damping_MF}) 
as a typical value.  
We solve the mean-field equations self-consistently at $T=0$ in the several system sizes 
for $N=L_x^2$ with $L_x=112$, 800, 1200, 1600, 1920, and 2240. 
Below we will show the results calculated 
on the lattice sites with $L_x=1200$ unless otherwise noted.   

\subsubsection{f-electron number per site and the characteristic energy}

\begin{figure}
\includegraphics[width=7.5cm]{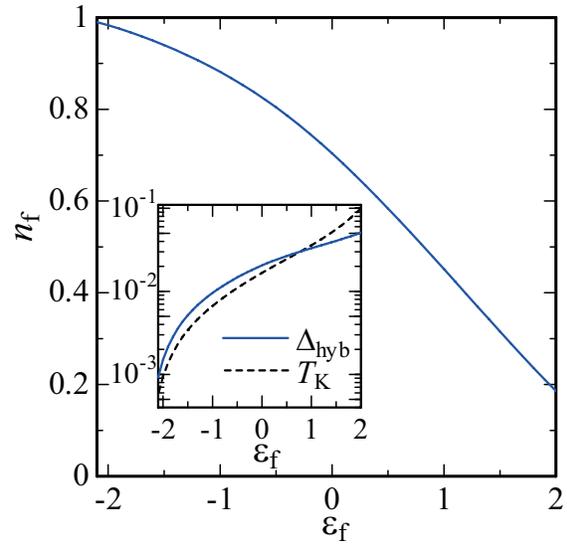}
\caption{
(Color online) The $\varepsilon_{\rm f}$ dependence of the f-electron number per site 
for $t=1$, $V=0.3$, and $U=\infty$ at $\bar{n}=7/4$ with $\Gamma=10^{-3}$ 
calculated in the $N=1200\times 1200$ lattice sites. 
Inset: The $\varepsilon_{\rm f}$ dependence of $\Delta_{\rm hyb}$ (solid line) 
and $T_{\rm K}$ (dashed line). 
}
\label{fig:nf_Ef}
\end{figure}

Figure~\ref{fig:nf_Ef} shows the $\varepsilon_{\rm f}$ dependence of 
the f-electron number per site, $n_{\rm f}$. 
As $\varepsilon_{\rm f}$ increases, the crossover from the Kondo regime with $n_{\rm f}\approx 1$ 
in the deep-$\varepsilon_{\rm f}$ region 
to the valence-fluctuation~\cite{VF} regime with $n_{\rm f}<1$ 
in the shallow-$\varepsilon_{\rm f}$ region 
occurs in the ground state.  

The characteristic energy scale of the present system, 
which is given by the hybridization gap 
$\Delta_{\rm hyb}$, is defined by the energy gap between the bottom of 
the upper hybridized band and the top of the lower hybridized band of quasiparticles:
\begin{eqnarray}
\Delta_{\rm hyb}\equiv \tilde{E}^{+}_{{\bf k}=(0,0)}-\tilde{E}^{-}_{{\bf k}=(\pi,\pi)}. 
\label{eq:gap_hyb}  
\end{eqnarray}
Since we consider the filling of $\bar{n}=7/4$ less than half filling, the Fermi level is located at the lower hybridized band. 
The Kondo temperature $T_{\rm K}$, which is the characteristic energy scale of 
the heavy-electron system, is defined as the energy difference between 
the renormalized f level and the Fermi level
$T_{\rm K}\equiv\tilde{\varepsilon}_{\rm f}-\mu$ 
in the present mean-field framework. 
The inset of Fig.~\ref{fig:nf_Ef} shows that 
$\Delta_{\rm hyb}$ (solid line) roughly corresponds to $T_{\rm K}$ (dashed line), 
both of which well scale for $\varepsilon_{\rm f}\lsim 0$. 

\begin{figure}
\includegraphics[width=7.5cm]{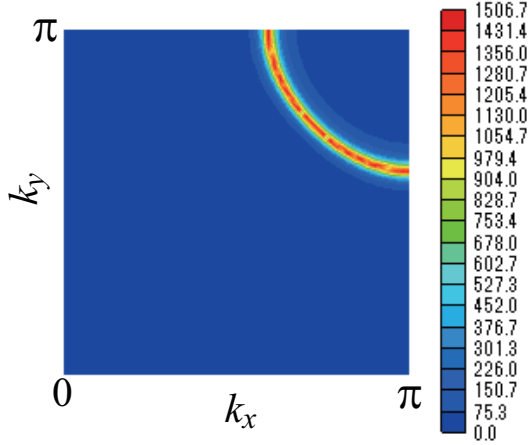}
\caption{
(Color online) The contour plot of the spectral function $A_{-}({\bf k},\mu)=-{\rm Im}[G_{\bf k}^{-{\rm R}}(\mu)]/\pi$ at $\bar{n}=7/4$ for $t=1$, $V=0.3$, $\varepsilon_{\rm f}=-4.0$, and $U=\infty$ 
with $\Gamma=10^{-3}$ 
calculated in the $N=112\times 112$ lattice sites.
}
\label{fig:n78_FS}
\end{figure}

To visualize the Fermi surface at $\bar{n}=7/4$, we plot the spectral function 
$A_{\alpha}({\bf k},\varepsilon)\equiv-\frac{1}{\pi}{\rm Im}G_{\bf k}^{\alpha{\rm R}}(\varepsilon)$ for $\alpha=-$ and $\varepsilon=\mu$ 
in Fig.~\ref{fig:n78_FS}. Here we show the contour plot for $\varepsilon_{\rm f}=-4.0$ as a typical case calculated in the $N=112\times 112$ lattice sites.  

\subsubsection{Diagonal conductivity}
\label{sec:sxx}

\begin{figure}
\includegraphics[width=7.5cm]{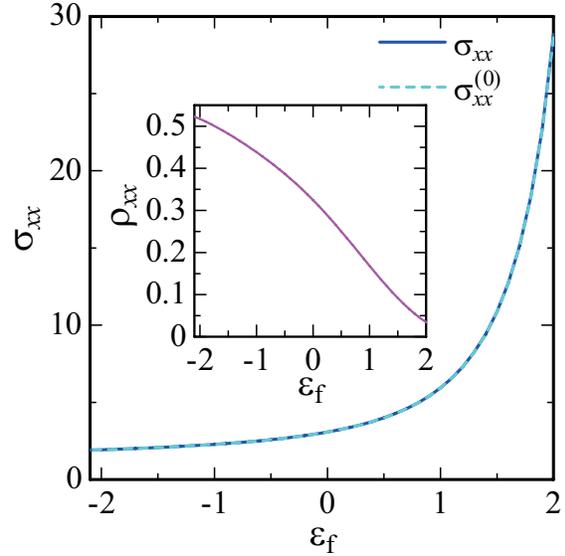}
\caption{
(Color online) The $\varepsilon_{\rm f}$ dependence of the conductivity $\sigma_{xx}$ 
(solid line) and $\sigma_{xx}^{(0)}$ (dashed line) 
for $t=1$, $V=0.3$, and $U=\infty$ at $\bar{n}=7/4$ with $\Gamma=10^{-3}$ 
is shown in the left axis, which is calculated in the $N=1200\times 1200$ lattice sites. 
We set $e=1$. Inset shows the $\varepsilon_{\rm f}$-level dependence of resistivity $\rho_{xx}=1/\sigma_{xx}$. 
}
\label{fig:sxx_Ef}
\end{figure}

By using the mean-field solutions, the conductivity is calculated on the basis of 
Eq.~(\ref{eq:sxx_tot}) at $T=0$. 
The $\varepsilon_{\rm f}$ dependence of $\sigma_{xx}$ is shown 
in Fig.~\ref{fig:sxx_Ef}. 
In the deep-$\varepsilon_{\rm f}$ region, 
as $\varepsilon_{\rm f}$ increases, 
$\sigma_{xx}$ shows a gradual increase which can be seen as almost constant behavior, 
while $\sigma_{xx}$ shows a sharp increase 
in the shallow-$\varepsilon_{\rm f}$ region for $\varepsilon_{\rm f}\gsim 0$. 
The inset shows the $\varepsilon_{\rm f}$-level dependence of the resistivity $\rho_{xx}=1/\sigma_{xx}$. 

To analyze the mechanism, 
we plot $\sigma_{xx}^{(0)}$ in Fig.~\ref{fig:sxx_Ef}, which is defined by
\begin{eqnarray}
\sigma_{xx}^{(0)}=\frac{2e^2}{(2\pi)^2}\sum_{{\bf k}={\bf k}_{\rm F}}
\frac{
\left(v_{{\bf k}x}^{\rm c0}\right)^2\left({\tilde{a}_{-,{\bf k}}^{\rm cc}}\right)^2
}{\left|\nabla \tilde{E}_{\bf k}^{-}\right|}\tilde{\tau}_{\bf k}^{-}\left|\Delta {\bf k}\right|,
\label{eq:sxx_2D_Gm0}
\end{eqnarray}
with $\tilde{\tau}_{\bf k}^{-}=\frac{1}{2\tilde{\Gamma}_{\bf k}^{-}}$. 
Here, the summation is taken over the Fermi wave vector ${\bf k}_{\rm F}$ 
and $|\Delta {\bf k}|$ is the length between each next ${\bf k}_{\rm F}$ point.  
This is the two-dimensional version of Eq.~(\ref{eq:sxx_iso_1}) in the lattice system. 
We see that the result almost coincides with $\sigma_{xx}$. 
This indicates that $\sigma_{xx}^{--}$ 
in Eq.~(\ref{eq:sxx_mm}) gives the dominant contribution to $\sigma_{xx}$ 
in Eq.~(\ref{eq:sxx_tot}) 
and the analysis with the small $\Gamma_{\bf k}^{-}$ in Eq.~(\ref{eq:sxx_Boltzmann}) is applicable 
to the parameter regime shown in Fig.~\ref{fig:sxx_Ef}. 
Since the velocity of the lower hybridized band $\tilde{\bf v}_{\bf k}^{--}\equiv\nabla\tilde{E}_{\bf k}^{-}$ is given by 
${\bf v}_{\bf k}^{--}={\bf v}_{\bf k}^{c0}\tilde{a}_{-,{\bf k}}^{\rm cc}$ 
[see Eq.~(\ref{eq:v_aa})], 
Eq.~(\ref{eq:sxx_2D_Gm0}) can be expressed as  
\begin{eqnarray} 
\sigma_{xx}^{(0)}=\frac{2e^2}{(2\pi)^2}\sum_{{\bf k}={\bf k}_{\rm F}}
\frac{
\left(v_{{\bf k}x}^{\rm c0}\right)^2{\tilde{a}_{-,{\bf k}}^{\rm cc}}
}{\left|{\bf v}_{\bf k}^{\rm c0}\right|2\tilde{\Gamma}_{\bf k}^{-}}\left|\Delta {\bf k}\right|. 
\label{eq:sxx0_2}
\end{eqnarray}
This implies that the ratio of the conduction-electron weight factor 
$\tilde{a}_{-,{\bf k}_{\rm F}}^{\rm cc}$ and the damping rate $\tilde{\Gamma}_{{\bf k}_{\rm F}}^{-}$ determines 
the behavior of $\sigma_{xx}^{(0)}$. 
This gives essentially the same form as Eq.~(\ref{eq:sxx_full}), 
which was formulated on the basis of the Fermi-liquid theory~\cite{Yamada}, as shown in Appendix. 
This indicates the validity of the present formalism. 
 
\begin{figure}
\includegraphics[width=7.5cm]{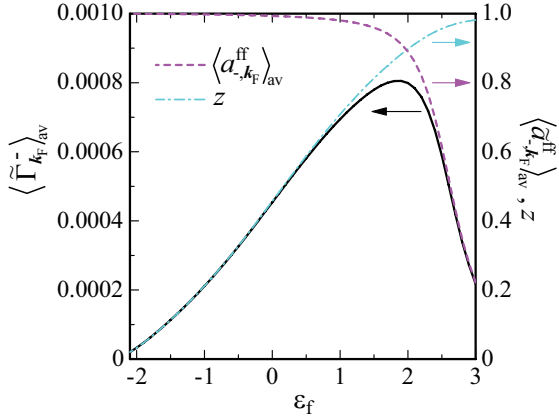}
\caption{
(Color online) The $\varepsilon_{\rm f}$ dependence of the f-electron weight factor 
$\langle \tilde{a}_{-,{\bf k}_{\rm F}}^{\rm ff}\rangle_{\rm av}$ 
(dashed line) and the renormalization factor $z$ (dash-dotted line)  
for $t=1$, $V=0.3$, and $U=\infty$ at $\bar{n}=7/4$ with $\Gamma=10^{-3}$ 
is shown in the right axis, which is calculated in the $N=1200\times 1200$ lattice sites.  
The $\varepsilon_{\rm f}$ dependence of the imaginary part of the self energy 
$\langle{\tilde{\Gamma}_{{\bf k}_{\rm F}}^{-}}\rangle_{\rm av}$ is also plotted (solid line) 
in the left axis. 
}
\label{fig:q_renm_aff_Ef}
\end{figure}

In order to clarify the $\varepsilon_{\rm f}$ dependence of $\tilde{\Gamma}_{\bf k}^{-}$ 
from Eq.~(\ref{eq:damping_MF}) 
in detail, we plot in Fig.~\ref{fig:q_renm_aff_Ef} 
the $\varepsilon_{\rm f}$ dependence of 
the renormalization factor ${z}$ and the f-electron weight factor 
which is averaged over the Fermi surface 
\begin{eqnarray}
\langle \tilde{a}_{-,{\bf k}_{\rm F}}^{\rm ff} \rangle_{\rm av}
\equiv\frac{1}{N_{{\bf k}_{\rm F}}}\sum_{{\bf k}={\bf k_{\rm F}}}
\tilde{a}_{-,{\bf k}}^{\rm ff} 
\label{eq:aff_2D}
\end{eqnarray}
with $N_{{\bf k}_{\rm F}}$ being the number of the ${\bf k}_{\rm F}$ points. 
As $\varepsilon_{\rm f}$ increases, 
$\langle \tilde{a}_{-,{\bf k}_{\rm F}}^{\rm ff} \rangle_{\rm av}$ 
is kept to be almost $1$ up to $\varepsilon_{\rm f}\sim 1$ and 
sharply decreases to zero for $\varepsilon_{\rm f}\gsim 1$. 

The renormalization factor ${z}$ approaches zero in the deep-$\varepsilon_{\rm f}$ limit 
due to strong correlation effect on f electrons with $n_{\rm f}\to 1$ (see Fig.~\ref{fig:nf_Ef}). 
As $\varepsilon_{\rm f}$ increases, ${z}$ increases gradually. 
The damping rate $\tilde{\Gamma}_{\bf k}^{-}$ given by the multiplication of 
$\tilde{a}_{-,{\bf k}}^{\rm ff}$ and ${z}$  
in Eq.~(\ref{eq:damping_MF}) is averaged over the Fermi surface 
\begin{eqnarray}
\langle\tilde{\Gamma}_{{\bf k}_{\rm F}}^{-}\rangle_{\rm av}
=\frac{1}{N_{{\bf k}_{\rm F}}}\sum_{{\bf k}={\bf k}_{\rm F}}\tilde{\Gamma}_{\bf k}^{-},
\label{eq:Gamma_av}
\end{eqnarray}
which 
shows a peak structure around $\varepsilon_{\rm f}\sim 2$ as shown in Fig.~\ref{fig:q_renm_aff_Ef}. 
An important result here is that $\langle\tilde{\Gamma}_{{\bf k}_{\rm F}}^{-}\rangle_{\rm av}$ is 
suppressed compared to $\Gamma=10^{-3}$ in all the $\varepsilon_{\rm f}$ region since 
in both the large-$\varepsilon_{\rm f}$ and small-$\varepsilon_{\rm f}$ limits 
$\tilde{\Gamma}_{{\bf k}_{\rm F}}^{-}$ approaches zero and the peak value is bounded by the small $\Gamma$.

\begin{figure}
\includegraphics[width=7.5cm]{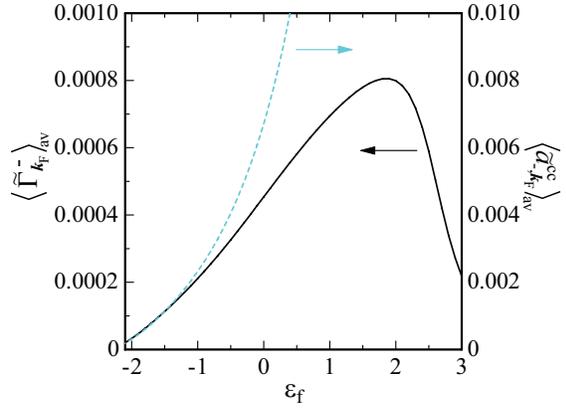}
\caption{
(Color online) The $\varepsilon_{\rm f}$ dependence of the conduction-electron weight factor 
$\langle \tilde{a}_{-,{\bf k}_{\rm F}}^{\rm cc}\rangle_{\rm av}$ 
(dashed line, right axis) and the imaginary part of the selfenergy 
$\langle{\tilde{\Gamma}_{{\bf k}_{\rm F}}^{-}}\rangle_{\rm av}$ (solid line, left axis)  
is shown, which is calculated 
for $t=1$, $V=0.3$, and $U=\infty$ at $\bar{n}=7/4$ with $\Gamma=10^{-3}$ 
in the $N=1200\times 1200$ lattice sites.  
}
\label{fig:ImS_acc_Ef}
\end{figure}

Figure~\ref{fig:ImS_acc_Ef} shows the $\varepsilon_{\rm f}$ dependence of 
the conduction-electron weight factor averaged over the Fermi surface 
\begin{eqnarray}
\langle \tilde{a}_{-,{\bf k}_{\rm F}}^{\rm cc} \rangle_{\rm av}
\equiv\frac{1}{N_{{\bf k}_{\rm F}}}\sum_{{\bf k}={\bf k_{\rm F}}}
\tilde{a}_{-,{\bf k}}^{\rm cc}. 
\label{eq:acc_2D}
\end{eqnarray}
In Fig.~\ref{fig:ImS_acc_Ef}, $\langle\tilde{\Gamma}_{{\bf k}_{\rm F}}^{-}\rangle_{\rm av}$ is also re-plotted 
for comparison. 
As $\varepsilon_{\rm f}$ increases, 
$\langle \tilde{a}_{-,{\bf k}_{\rm F}}^{\rm cc} \rangle_{\rm av}$ 
increases gradually in the deep-$\varepsilon_{\rm f}$ region, while 
it shows a sharp increase around $\varepsilon_{\rm f}\sim 1$. 
The gradual increase in 
$\langle \tilde{a}_{-,{\bf k}_{\rm F}}^{\rm cc} \rangle_{\rm av}$ 
and 
$\langle\tilde{\Gamma}_{{\bf k}_{\rm F}}^{-}\rangle_{\rm av}$
in the deep-$\varepsilon_{\rm f}$ region gives rise to cancellation of the effect 
of the mass renormalization~\cite{Yamada} in Eq.~(\ref{eq:sxx0_2}), which causes 
the almost constant $\varepsilon_{\rm f}$ dependence of $\sigma_{xx}^{(0)}$. 
However, as $\varepsilon_{\rm f}$ increases to reach the shallow-$\varepsilon_{\rm f}$ region, 
$\varepsilon_{\rm f}\gsim 0$, i.e., so-called the ``valence-fluctuation" regime, 
the cancellation does not work, where $a_{-,{\bf k}_{\rm F}}^{\rm cc}$ increases sharply 
while $\tilde{\Gamma}_{{\bf k}_{\rm F}}^{-}$ remains small. 
This imbalance is the reason why $\sigma_{xx}$ shows a sharp increase in the valence-fluctuation regime for $\varepsilon_{\rm f}\gsim 0$ in Fig.~\ref{fig:sxx_Ef}. 
This gives a natural explanation for the pressure dependence of the residual resistivity 
frequently observed in the Ce-based compounds and Yb-based compounds.    
The pressure dependence of the conductivity will be discussed in detail 
in Sect.~\ref{sec:hyb_dep}. 

The above result is obtained by using constant $\Gamma$ in Eq.~(\ref{eq:Gamma_def}). 
As noted below Eq.~(\ref{eq:S_imp}), $\langle a^{\rm ff}_{-,{{\bf k}_{\rm F}}}\rangle_{\rm av}N^{*}(\mu)$ can be expressed essentially by the bare quantities, which is on the order of $O(\pi V^2N_{\rm cF})^{-1}$. Hence, $\Gamma$ defined in Eq.~(\ref{eq:Gamma_def}) has only weak-$\varepsilon_{\rm f}$ dependence. However, as shown in Fig.~\ref{fig:q_renm_aff_Ef}, the quantities related to renormalization factor, $z$ and $\tilde{a}^{\rm ff}_{-,{\bf k}}$ in Eq.~(\ref{eq:damping_MF}), have strong $\varepsilon_{\rm f}$ dependence, which give the main contribution to the remarkable change of $\sigma_{xx}$ when $\varepsilon_{\rm f}$ varies from the Kondo regime to the valence-fluctuation regime. Hence, present treatment using constant $\Gamma$ is considered to capture the essence of the transport phenomena. 
As for the hybridization dependence, we have also performed the calculations of the $\varepsilon_{\rm f}$ dependence of $\sigma_{xx}$ by inputting several values of $\Gamma$ and confirmed that the main conclusion above does not change as far as the renormalized damping rate is far smaller than the hybridization gap.  
The $\Gamma$ dependence and the $V$ dependence will be discussed in Sect. 4.2.6 and Sect. 4.2.7, respectively.  

\subsubsection{Hall conductivity}
\label{sec:sxyH}

\begin{figure}
\includegraphics[width=7.5cm]{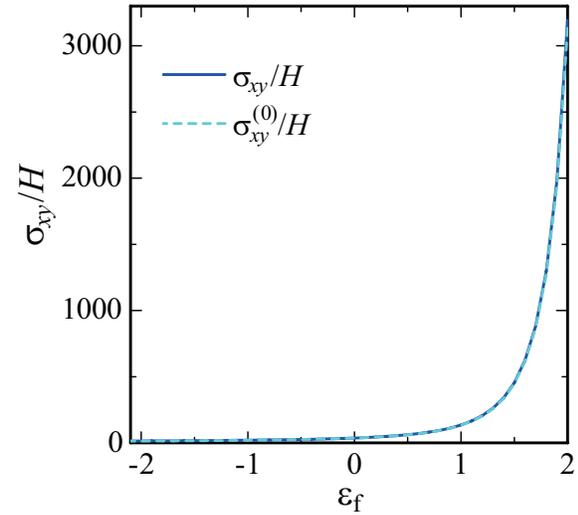}
\caption{
(Color online) The $\varepsilon_{\rm f}$ dependence of the conductivity $\sigma_{xy}/H$ 
(solid line) and $\sigma_{xy}^{(0)}/H$ (dashed line) 
for $t=1$, $V=0.3$, and $U=\infty$ at $\bar{n}=7/4$ with $\Gamma=10^{-3}$ 
is shown in the left axis, which is calculated in the $N=1200\times 1200$ lattice sites. 
We set $e=1$.  
}
\label{fig:sxyH_Ef}
\end{figure}

The $\varepsilon_{\rm f}$ dependence of the Hall conductivity is shown in Fig.~\ref{fig:sxyH_Ef}. 
The Hall conductivity $\sigma_{xy}/H$ is calculated by using Eq.~(\ref{eq:sxy}) 
at $T=0$. 
As $\varepsilon_{\rm f}$ increases, in the deep-$\varepsilon_{\rm f}$ region 
$\sigma_{xy}/H$ gradually increases, which can be seen 
as almost constant behavior, while it shows a sharp increase in the shallow-$\varepsilon_{\rm f}$ region 
for $\varepsilon_{\rm f}\gsim 0$. 

To analyze the mechanism, we plot 
$\sigma_{xy}^{(0)}/H$ in Fig.~\ref{fig:sxyH_Ef}, which is defined by
\begin{eqnarray}
\frac{\sigma_{xy}^{(0)}}{H}&=&-\frac{2e^3}{(2\pi)^2}
\sum_{{\bf k}={\bf k}_{\rm F}}
\frac{
\left(v_{{\bf k}x}^{\rm c0}\right)^2
\left(
\frac{\partial v_{{\bf k}y}^{{\rm c}0}}{\partial k_y}
\right)
\left({\tilde{a}_{-,{\bf k}}^{\rm cc}}\right)^3
}
{\left|\nabla \tilde{E}_{\bf k}^{-}\right|}
\left({\tilde{\tau}_{\bf k}^{-}}\right)^2
\left|\Delta {\bf k}\right| 
\label{eq:sxy_2D_Gm0}
\end{eqnarray}
with the same notation as Eq.~(\ref{eq:sxx_2D_Gm0}).
This is the two-dimensional version of Eq.~(\ref{eq:sxy_ds}) in the lattice system. 
We see that the result almost coincides with $\sigma_{xy}/H$. 
This indicates that 
$\sigma_{xy}^{--}/H$ in Eq.~(\ref{eq:sxy_1}) dominantly contributes to $\sigma_{xy}/H$ 
in Eq.~(\ref{eq:sxy})  
and the analysis by the small $\tilde{\Gamma}_{\bf k}^{-}=\frac{1}{2\tilde{\tau}_{\bf k}^{-}}$  
in Eq.~(\ref{eq:sxy_T0}) is applicable to the parameter regime shown in Fig.~\ref{fig:sxyH_Ef}. 
By using 
the velocity of the lower hybridized band, 
$\tilde{\bf v}_{\bf k}^{--}=\nabla \tilde{E}_{\bf k}^{-}={\bf v}_{\bf k}^{\rm c0}a_{-,{\bf k}}^{\rm cc}$, 
Eq.~(\ref{eq:sxy_2D_Gm0}) can be expressed as
\begin{eqnarray}
\frac{\sigma_{xy}^{(0)}}{H}&=&-\frac{2e^3}{(2\pi)^2}
\sum_{{\bf k}={\bf k}_{\rm F}}
\frac{
\left(v_{{\bf k}x}^{\rm c0}\right)^2
\left(
\frac{\partial v_{{\bf k}y}^{{\rm c}0}}{\partial k_y}
\right)
\left({\tilde{a}_{-,{\bf k}}^{\rm cc}}\right)^2
}
{\left|{\bf v}_{\bf k}^{\rm c0}\right|
\left(2\tilde{\Gamma}_{\bf k}^{-}\right)^2}
\left|\Delta {\bf k}\right|.  
\label{eq:sxy0_2}
\end{eqnarray}
In the right hand side, 
$(\tilde{a}_{-,{\bf k}_{\rm F}}^{\rm cc}/\tilde{\Gamma}_{{\bf k}_{\rm F}}^{-})^2$ appears, 
which implies that the ratio 
$\tilde{a}_{-,{\bf k}_{\rm F}}^{\rm cc}/\tilde{\Gamma}_{{\bf k}_{\rm F}}^{-}$ determines the behavior of $\sigma_{xy}^{(0)}/H$. 
As shown in Fig.~\ref{fig:ImS_acc_Ef}, in the deep-$\varepsilon_{\rm f}$ region, 
$\tilde{a}_{-,{\bf k}_{\rm F}}^{\rm cc}/\tilde{\Gamma}_{{\bf k}_{\rm F}}^{-}$ increases gradually, 
while it shows a sharp increase in the shallow-$\varepsilon_{\rm f}$ region for 
$\varepsilon_{\rm f}\gsim 0$. 
Namely, cancellation of the effect of the mass enhancement, i.e., 
$\tilde{a}_{-,{\bf k}_{\rm F}}^{\rm cc}/\tilde{\Gamma}_{{\bf k}_{\rm F}}^{-}\approx 1$, 
causes the almost constant 
behavior of $\sigma_{xy}/H$ in the deep-$\varepsilon_{\rm f}$ region and 
the imbalance, i.e., 
$\tilde{a}_{-,{\bf k}_{\rm F}}^{\rm cc}/\tilde{\Gamma}_{{\bf k}_{\rm F}}^{-}\gg 1$, 
makes the sharp increase in $\sigma_{xy}/H$ in the valence-fluctuation regime 
for $\varepsilon_{\rm f}\gsim 0$
in Fig.~\ref{fig:sxyH_Ef}. 
 
\subsubsection{Hall coefficient}
\label{sec:RH}

\begin{figure}
\includegraphics[width=7.5cm]{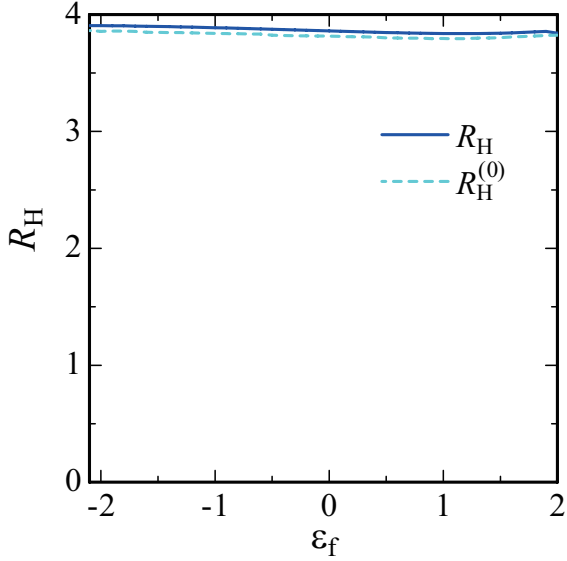}
\caption{
(Color online) The $\varepsilon_{\rm f}$ dependence of the conductivity $R_{\rm H}$ 
(solid line) and $R_{\rm H}^{(0)}$ (dashed line) 
for $t=1$, $V=0.3$, and $U=\infty$ at $\bar{n}=7/4$ with $\Gamma=10^{-3}$ 
is shown in the left axis, which is calculated in the $N=1200\times 1200$ lattice sites. 
We set $e=1$.  
}
\label{fig:RH_Ef}
\end{figure}

The $\varepsilon_{\rm f}$ dependence of the Hall coefficient is shown in Fig.~\ref{fig:RH_Ef}. 
Here we set $e=1$. 
The Hall coefficient 
$R_{\rm H}=\sigma_{xy}/(H\sigma_{xx}^2)$ (solid line) calculated by using 
Eqs.~(\ref{eq:sxx_tot}) and (\ref{eq:sxy}) shows a 
slight decrease as a function of $\varepsilon_{\rm f}$, exhibiting   
almost constant behavior. 
Namely, the sharp increase in $\sigma_{xx}$ and $\sigma_{xy}/H$ in the valence-fluctuation 
regime for $\varepsilon_{\rm f}\gsim 0$ in Fig.~\ref{fig:sxx_Ef} and Fig.~\ref{fig:sxyH_Ef}, 
respectively, cancel out each other in $R_{\rm H}$. 
To analyze this cancellation which occurs even in the valence-fluctuation regime, 
let us plot 
$R_{\rm H}^{(0)}=\sigma_{xy}^{(0)}/(H{\sigma_{xx}^{(0)}}^2)$ by the dashed line  
calculated by using Eqs.~(\ref{eq:sxx_2D_Gm0}) and (\ref{eq:sxy_2D_Gm0}) 
in Fig.~\ref{fig:RH_Ef}. 
We see that 
$R_{\rm H}^{(0)}$ also exhibits almost constant behavior. 
This is understood from expressions Eq.~(\ref{eq:sxx_2D_Gm0}) and Eq.~(\ref{eq:sxy_2D_Gm0}) 
since the factors of $\tilde{a}_{-,{\bf k}_{\rm F}}^{\rm cc}/\tilde{\Gamma}_{{\bf k}_{\rm F}}^{-}$ 
which appear in both $\sigma_{xx}^{(0)}$ and $\sigma_{xy}^{(0)}/H$ 
cancel out in $R_{\rm H}^{(0)}$.  
The close agreement between $R_{\rm H}$ and $R_{\rm H}^{(0)}$ indicates that 
the cancellation of the factors of 
$\tilde{a}_{-,{\bf k}_{\rm F}}^{\rm cc}/\tilde{\Gamma}_{{\bf k}_{\rm F}}^{-}$ actually 
occurs in $R_{\rm H}$, as was shown for the isotropic free-electrons in Eq.~(\ref{eq:1ne}). 
However, in Fig.~\ref{fig:RH_Ef}, 
the sign in $R_{\rm H}$ is positive and the magnitude is not 
expressed by the total filling as 
$1/({\bar{n}e})=4/(7e)$, which are in contrast to the result to Eq.~(\ref{eq:1ne}). 
These points will be analyzed below. 

\begin{figure}
\includegraphics[width=7.5cm]{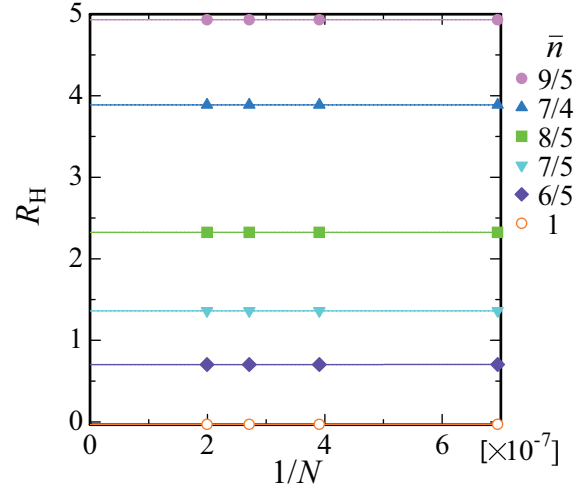}
\caption{
(Color online) The system-size dependence of the Hall coefficient $R_{\rm H}$  
for $t=1$, $V=0.3$, $\varepsilon_{\rm f}=-4.0$, and $U=\infty$ with $\Gamma=10^{-3}$ 
at $\bar{n}=1$ (open circle), $6/5$ (filled diamond), $7/5$ (filled inverted triangle), 
$8/5$ (filled square), $7/4$ (filled triangle), and $9/5$ (filled circle). 
We set $e=1$. 
}
\label{fig:RH_N}
\end{figure}
 
To figure out the filling dependence of $R_{\rm H}$, 
we calculate $\sigma_{xx}$ and $\sigma_{xy}/H$ for $1\le\bar{n}<2$ 
plausible to the heavy-electron state 
in several system sizes 
and extrapolate $R_{\rm H}$ to the bulk limit, $N=L_x^2\to\infty$.
Figure~\ref{fig:RH_N} shows $R_{\rm H}$ vs. $1/N$ for $\varepsilon_{\rm f}=-4.0$ 
at $\bar{n}=1$, 6/5, 7/5, 8/5, 7/4, and 9/5. 
The system sizes used for the extrapolation are 
$N=L_x^2$ with 
$L_x=1200$, 1600, 1920, and 2240.

\begin{figure}
\includegraphics[width=7.5cm]{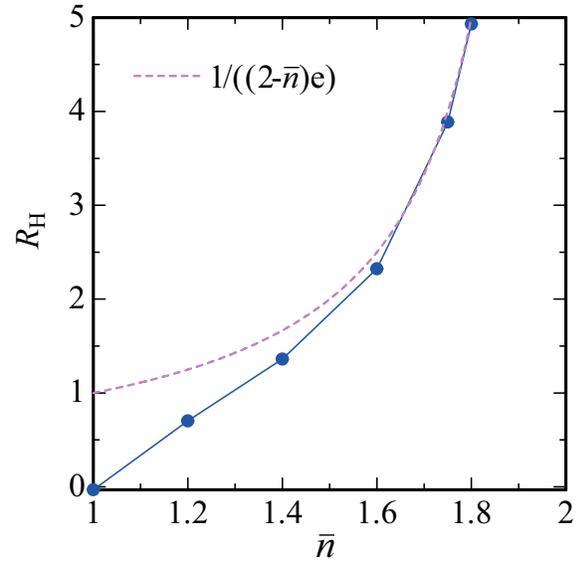}
\caption{
(Color online) The filling dependence of the Hall coefficient $R_{\rm H}$  
for $t=1$, $V=0.3$, $\varepsilon_{\rm f}=-4.0$, and $U=\infty$ with $\Gamma=10^{-3}$ 
in the bulk limit $(N\to\infty)$. 
The dashed line represents $1/((2-\bar{n})e)$. We set $e=1$.   
}
\label{fig:RH_bulk_N}
\end{figure}
  
The $\bar{n}$ dependence of $R_{\rm H}$ in the bulk limit is shown in Fig.~\ref{fig:RH_bulk_N}. 
Note that the error bar by the least-square fit done for the system-size extrapolation 
is attached to each filled circle. 
The error bars are within the symbol sizes and invisible, 
indicating that the system size dependence does not matter in the $N\ge 1200$ lattice sites. 
Here, we also plot the Hall coefficient expressed by the hole density as 
$R_{\rm H}^{\rm hole}=\frac{1}{\bar{n}_{\rm hole}e}$ 
with $\bar{n}_{\rm hole}\equiv 2-\bar{n}$ by a dashed line (Note that we set $e=1$). 
We can see that $R_{\rm H}$ approaches $R_{\rm H}^{\rm hole}$ as $\bar{n}$ approaches 
half filling, $\bar{n}=2$. 
However, $R_{\rm H}$ approaches zero as $\bar{n}$ approaches $1$, which shows 
a clear deviation from the dashed line.  
These results indicate that 
near the quarter filling, i.e., for $1\le\bar{n}\lsim 1.6$, 
$R_{\rm H}$ is not expressed simply by the hole density as 
$\frac{1}{\bar{n}_{\rm hole}e}$. 

\subsubsection{Curvature of the Fermi surface and Hall conductivity and Hall coefficient}
\label{sec:sxx_RH}

To understand the reason why $R_{\rm H}$ does not follow the simple relation $R_{\rm H}=1/({\bar n}_{\rm hole}e)$, we analyze $\sigma_{xy}/H$ from the viewpoint of the curvature 
of the Fermi surface. In the single-band system constituted of a single orbital, it has been 
shown that $\sigma_{xy}/H$ can be expressed by the angle between $\vec{v}_{\bf k}$ 
and the $k_x$ axis within the Boltzmann transport theory~\cite{Tsuji,Ziman,Ong}
and the theory considering the vertex corrections\cite{Kontani1999}. 
In the present system, there exist two orbitals of f and conduction electrons, 
which form the lower and upper hybridized bands. 
Since the Fermi level is located at the lower hybridized band, 
the system is regarded as the single-band system at $T=0$ in the small-$\tilde{\Gamma}_{\bf k}^{-}$ limit. 

At sufficiently low temperatures in the small damping rate, 
$\sigma_{xy}^{--}/H$ in Eq.~(\ref{eq:sxy_1}) dominantly contributes to $\sigma_{xy}/H$ 
in Eq.~(\ref{eq:sxy}). 
Let us write Eq.~(\ref{eq:sxy_simple}) in the original form like Eq.~(\ref{eq:sxy}) as 
\begin{eqnarray}
\sigma_{xy}/H&=&
-\frac{e^3}{2V_{0}}\sum_{{\bf k}\sigma}
\left(
-\frac{\partial f(\tilde{E}_{\bf k}^{-})}{\partial \tilde{E}_{\bf k}^{-}}
\right)
\nonumber
\\
& &
\times
\tilde{v}_{{\bf k}x}^{-}
\left[
\tilde{v}_{{\bf k}x}^{-}
\left(
\frac{\partial \tilde{v}_{{\bf k}y}^{-}}{\partial k_{y}}
\right)
-\tilde{v}_{{\bf k}y}^{-}
\left(
\frac{\partial \tilde{v}_{{\bf k}x}^{-}}{\partial k_{y}}
\right)
\right]
\frac{2}{4(\tilde{\Gamma}_{\bf k}^{-})^2}. 
\label{eq:sxy_simple3}
\end{eqnarray}
Here, by Eq.~(\ref{eq:v_aa}),  
the velocity of the lower hybridized band ${\bf v}_{\bf k}^{--}=\nabla\tilde{E}_{\bf k}^{-}$  
in the present system for Eq.~(\ref{eq:PAM_MF}) is written as 
\begin{eqnarray}
{\bf v}_{\bf k}^{--}=
{\bf v}_{\bf k}^{\rm c0}\tilde{a}_{-,{\bf k}}^{\rm cc}
\equiv\tilde{\bf v}_{\bf k}^{-}. 
\label{eq:vmband}
\end{eqnarray}
Now we apply the formalism shown in Ref.~\citen{Kontani1999} to Eq.~(\ref{eq:sxy_simple3}). 
Here we describe it up to Eq.~(\ref{eq:sxy_simple6}) below  
as the self-contained explanation although it was originally published 
in Ref.~\citen{Kontani1999} [see Eq.~(22) in Ref.~\citen{Kontani1999}]. 

For the subsequent discussion, we rewrite Eq.~(\ref{eq:sxy_simple3}) as follows
\begin{eqnarray}
\sigma_{xy}/H
=
-\frac{e^3}{2V_{0}}\sum_{{\bf k}\sigma}
\left(
-\frac{\partial f(\tilde{E}_{\bf k}^{-})}{\partial \tilde{E}_{\bf k}^{-}}
\right)
A_{xy}^{-}({\bf k})
\frac{2}{4(\tilde{\Gamma}_{\bf k}^{-})^2}, 
\label{eq:sxy_simple4}
\end{eqnarray}
by introducing 
\begin{eqnarray}
A_{xy}^{-}({\bf k})\equiv
\tilde{v}_{{\bf k}x}^{-}
\left[
\tilde{v}_{{\bf k}x}^{-}
\left(
\frac{\partial \tilde{v}_{{\bf k}y}^{-}}{\partial k_{y}}
\right)
-\tilde{v}_{{\bf k}y}^{-}
\left(
\frac{\partial \tilde{v}_{{\bf k}x}^{-}}{\partial k_{y}}
\right)
\right].
\end{eqnarray}
As shown in Ref.~\citen{Kohno} [see Eq.~(3.21) in Ref.~\citen{Kohno}],   
this can be rewritten in a simpler form as 
\begin{eqnarray}
\sigma_{xy}/H
=
-\frac{e^3}{4V_{0}}\sum_{{\bf k}\sigma}
\left(
-\frac{\partial f(\tilde{E}_{\bf k}^{-})}{\partial \tilde{E}_{\bf k}^{-}}
\right)
A_{\rm s}^{-}({\bf k})
\frac{2}{4(\tilde{\Gamma}_{\bf k}^{-})^2}, 
\label{eq:sxy_simple5}
\end{eqnarray}
where $A_{\rm s}({\bf k})$ is defined as
\begin{eqnarray}
A_{\rm s}^{-}({\bf k})&=&A_{xy}^{-}({\bf k})+A_{yx}^{-}({\bf k}).
\end{eqnarray}
This can be expressed as~\cite{Kontani1999}
\begin{eqnarray}
A_{\rm s}^{-}({\bf k})&=&
\tilde{v}_{{\bf k}x}^{-}(\vec{e}_{z}\times\tilde{\bf v}_{\bf k}^{-})
\nabla\tilde{v}_{{\bf k}y}^{-}
-\tilde{v}_{{\bf k}y}^{-}(\vec{e}_{z}\times\tilde{\bf v}_{\bf k}^{-})
\nabla\tilde{v}_{{\bf k}x}^{-}, 
\\
&=&\left|\tilde{\bf v}_{\bf k}^{-}\right|
\left(
\tilde{\bf v}_{\bf k}^{-}\times\frac{\partial}{\partial k_{\parallel}}\tilde{\bf v}_{\bf k}^{-}
\right)_{z}, 
\\
&=&\left|\tilde{\bf v}_{\bf k}^{-}\right|
\cdot
\left|\tilde{\bf v}_{\bf k}^{-}\right|^2
\left(\frac{d\theta_{\tilde{v}^{-}}({\bf k})}{dk_{\parallel}}\right), 
\label{eq:theta}
\end{eqnarray}
where $k_{\parallel}$ is the component of $\vec{k}$ along the vector 
$\vec{e}_{\parallel}({\bf k})=(\vec{e}_{z}\times\vec{\tilde{v}}_{\bf k}^{-})/\left|\tilde{\bf v}_{\bf k}^{-}\right|$, and tangential to the Fermi surface at $\bf k$ since $\tilde{\bf v}_{\bf k}^{-}$ is 
perpendicular to the Fermi surface. 
In Eq.~(\ref{eq:theta}), $\theta_{\tilde{v}^{-}({\bf k})}$ is the angle between 
$\tilde{\bf v}_{\bf k}^{-}$ and the $k_{x}$ axis. 

By applying the similar derivation used in Eq.~(\ref{eq:sxy_ds}) to Eq.~(\ref{eq:sxy_simple5}), 
$\sum_{\bf k}$ can be expressed by the line integral along the Fermi surface 
in the present two-dimensional system. 
Then we have
\begin{eqnarray}
\sigma_{xy}/H
=
-\frac{e^3}{4}
\frac{2}{(2\pi)^2}
\oint_{\rm FS}dk_{\parallel} 
\left|\tilde{\bf v}_{\bf k}^{-}\right|^2
\left(\frac{d\theta_{\tilde{v}^{-}}({\bf k})}{dk_{\parallel}}\right)
\frac{2}{4(\tilde{\Gamma}_{\bf k}^{-})^2},   
\label{eq:sxy_simple6}
\end{eqnarray}
where the $\bf k$ point moves counterclockwise along the Fermi surface in this line integral. 
Note that by Eq.~(\ref{eq:vmband}) the mass renormalization factors appear as 
$(\tilde{a}_{-,{\bf k}}^{\rm cc}/\tilde{\Gamma}_{\bf k}^{-})^2$ in the integrand, which cancel out 
in the deep-$\varepsilon_{\rm f}$ region as noted in Eq.~(\ref{eq:sxy0_2}). 

\begin{figure}
\includegraphics[width=8cm]{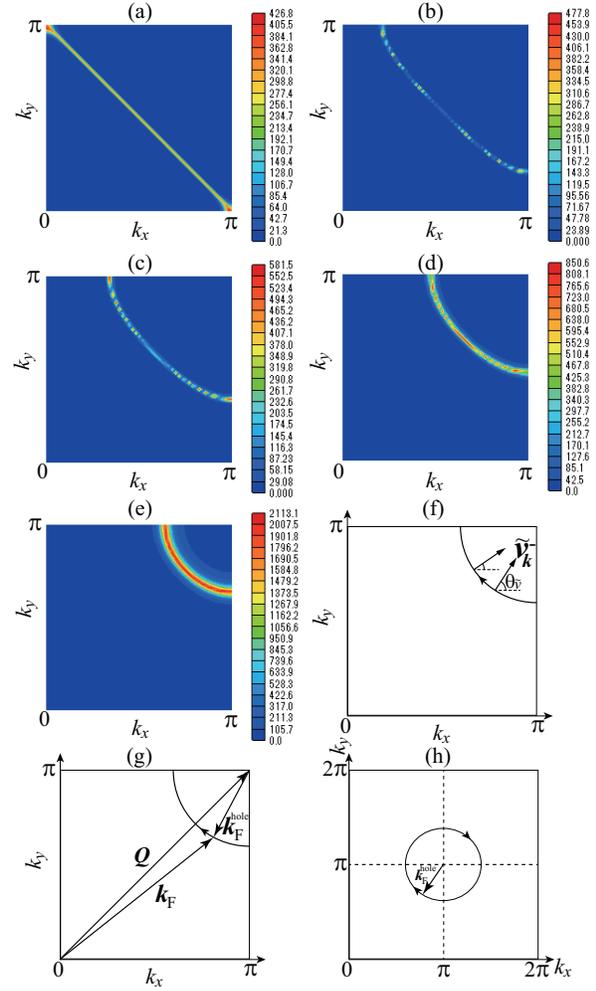}
\caption{(Color online) 
The contour plot of the spectral function $A_{-}({\bf k},\mu)=-{\rm Im}[G_{\bf k}^{-{\rm R}}(\mu)]/\pi$ at (a) $\bar{n}=1$, (b) $6/5$, (c) $7/5$, (d) $8/5$, and (e) $9/5$  
for $t=1$, $V=0.3$, $\varepsilon_{\rm f}=-4.0$, and $U=\infty$ 
with $\Gamma=10^{-3}$ 
calculated in the $N=112\times 112$ lattice sites. 
(f) The velocity of the lower hybridized band $\tilde{\bf v}_{\bf k}$ on the 
Fermi surface ${\bf k}={\bf k}_{\rm F}$. 
(g) The relation among ${\bf k}_{\rm F}$, ${\bf k}_{\rm F}^{\rm hole}$, and $\bf Q=(\pi,\pi)$,  
${\bf k}_{\rm F}={\bf Q}+{\bf k}^{\rm hole}_{\rm F}$. 
(h) The path for the line integral Eq.~(\ref{eq:sxy_simple7}) along the Fermi surface. 
}
\label{fig:FS_n_all}
\end{figure}

To visualize the Fermi surface, we plot the spectral function 
$A_{\alpha}({\bf k},\varepsilon)\equiv-{\rm Im}G_{\bf k}^{\alpha{\rm R}}(\varepsilon)/\pi$ 
for $\alpha=-$ and $\varepsilon=\mu$ 
at (a) $\bar{n}=1$, (b) 6/5, (c) 7/5, (d) 8/5, and (e) 9/5 in Fig.~\ref{fig:FS_n_all}. 
Here, set of parameters $(t=1, V=0.3, \varepsilon_{\rm f}=-4.0,$ and $U=\infty$ with $\Gamma=10^{-3})$ 
is the same as those in Fig.~\ref{fig:RH_N} and the results calculated in the $N=112 \times 112$ 
lattice sites are shown.  

At quarter filling, $\bar{n}=1$, we see that $\theta_{\tilde{v}_{\bf k}^{-}}({\bf k})$ 
for example on the first quadrant does not change, 
since $\tilde{\bf v}_{{\bf k}_{\rm F}}^{-}$ is perpendicular to the Fermi surface, 
which makes $\theta_{\tilde{v}^{-}}({\bf k}_{\rm F})$ be kept to be $\pi/4$.   
This gives rise to $d\theta_{\tilde{v}^{-}}({\bf k})/dk_{\parallel}=0$ in Eq.~(\ref{eq:sxy_simple6}). 
Hence, it turns out that $\sigma_{xy}/H$ and $R_{\rm H}$ as well become zero at $\bar{n}=1$ 
at least for the small-$\tilde{\Gamma}_{\bf k}^{-}$ limit. 
Actually, $R_{\rm H}$ at $\bar{n}=1$ in Figs.~\ref{fig:RH_N} and \ref{fig:RH_bulk_N}
is shown to be almost zero although the spectral function is broaden near ${\bf k}=(0,\pi)$ 
and $(\pi,0)$ in Fig.~\ref{fig:FS_n_all}(a) 
because of the finite damping rate $\tilde{\Gamma}_{\bf k}^{-}$.  

The reason why the sign of $\sigma_{xy}/H$ and the resultant $R_{\rm H}$ become positive 
for $1<\bar{n}<2$  
can be also understood from expression Eq.~(\ref{eq:sxy_simple6}). 
As illustrated in Fig.~\ref{fig:FS_n_all}(f), the angle $\theta_{\tilde{v}^{-}}({\bf k})$ 
becomes smaller as $k_{\parallel}$ moves along the Fermi surface. 
Namely, $d\theta_{\tilde{v}^{-}}({\bf k})/dk_{\parallel}<0$ in Eq.~(\ref{eq:sxy_simple6}) 
makes the sign of $\sigma_{xy}/H$ be positive, and hence the positive Hall coefficient 
appears, $R_{\rm H}>0$. 

Furthermore, the reason why $R_{\rm H}$ approaches $1/((2-\bar{n})e)$ as $\bar{n}$ approaches 
half filling, 2, in Fig.~\ref{fig:RH_bulk_N} 
can be also understood on the basis of Eq.~(\ref{eq:sxy_simple6}).
As $\bar{n}$ approaches $\bar{n}=2$, the form of the Fermi surface for holes approaches the circle 
around ${\bf k}=(\pi,\pi)$ as shown in Fig.~\ref{fig:FS_n_all}(e). 
Hence, it is convenient to introduce the variable 
transformation ${\bf k}_{\rm F}={\bf Q}+{\bf k}^{\rm hole}_{\rm F}$ 
with a constant shift ${\bf Q}=(\pi,\pi)$ 
in Eq.~(\ref{eq:sxy_simple6}), as shown in Fig.~\ref{fig:FS_n_all}(g). 
Then, Eq.~(\ref{eq:sxy_simple6}) is expressed as
\begin{eqnarray}
\sigma_{xy}/H
=
-\frac{e^3}{4}
\frac{2}{(2\pi)^2}
\oint_{\rm FS}dk_{\parallel}^{\rm hole} 
\left|\tilde{\bf v}_{\bf k}^{-}\right|^2
\left(\frac{d\theta_{\tilde{v}^{-}}({\bf k})}{dk_{\parallel}^{\rm hole}}\right)
\frac{2}{4(\tilde{\Gamma}_{\bf k}^{-})^2}.   
\label{eq:sxy_simple7}
\end{eqnarray}
When the Fermi surface is a circle, the integration can be easily performed 
as follows:
Since the line integral in Fig.~\ref{fig:FS_n_all}(h) is performed clockwise, the negative sign 
appears as 
$d\theta_{\tilde{v}^{-}}({\bf k})/dk_{\parallel}^{\rm hole}=-1/k_{\rm F}^{\rm hole}$. 
By using $\oint_{\rm FS}dk_{\parallel}^{\rm hole}=2\pi k_{\rm F}$, 
we obtain 
\begin{eqnarray}
\sigma_{xy}/H
=
\frac{e^3}{4}
\frac{1}{(2\pi)^2}
2\pi k_{\rm F}^{\rm hole} 
\left(\tilde{v}_{k_{\rm F}^{\rm hole}}^{-}\right)^2
\left(\frac{1}{k_{\rm F}^{\rm hole}}\right)
\frac{1}{(\tilde{\Gamma}_{{k}_{\rm F}}^{-})^2}.   
\label{eq:sxy_simple8}
\end{eqnarray}
From Eq.~(\ref{eq:sxx_iso_1}) applied to the two-dimensional system, 
it can be shown that 
$\sigma_{xx}$ for the hole Fermi surface with a circle shape is expressed as
\begin{eqnarray}
\sigma_{xx}=\frac{e^2}{2\pi}k_{\rm F}^{\rm hole}\tilde{v}_{k_{\rm F}}^{-}
\frac{1}{2\tilde{\Gamma}_{{k}_{\rm F}}^{-}}.
\label{eq:sxx_simple1}
\end{eqnarray}
By Eqs.~(\ref{eq:sxy_simple8}) and (\ref{eq:sxx_simple1}), 
the Hall coefficient is obtained as 
\begin{eqnarray}
R_{\rm H}=\frac{\sigma_{xy}}{H\sigma_{xx}^2}=\frac{1}{\bar{n}_{\rm hole}e}, 
\label{eq:RH_hole} 
\end{eqnarray}
where the hole density is given by $\bar{n}_{\rm hole}=({k_{\rm F}^{\rm hole}})^2/(2\pi)$ 
in the two-dimensional system. 
Then it is understandable that $R_{\rm H}$ is expressed by the hole density as 
$1/(\bar{n}_{\rm hole}e)$ 
with $\bar{n}_{\rm hole}\equiv 2-\bar{n}$ 
as $\bar{n}$ approaches half filling, ${\bar n}=2$, in Fig.~\ref{fig:RH_bulk_N}. 

\subsubsection{Damping-rate dependence}
\label{sec:Gm_dep}

So far, we have presented the results for the damping rate $\Gamma=10^{-3}$ [see Eq.~(\ref {eq:Gamma_def})] as a typical case. In this Subsect., we discuss the $\Gamma$ dependence. 

\begin{figure}
\includegraphics[width=7.5cm]{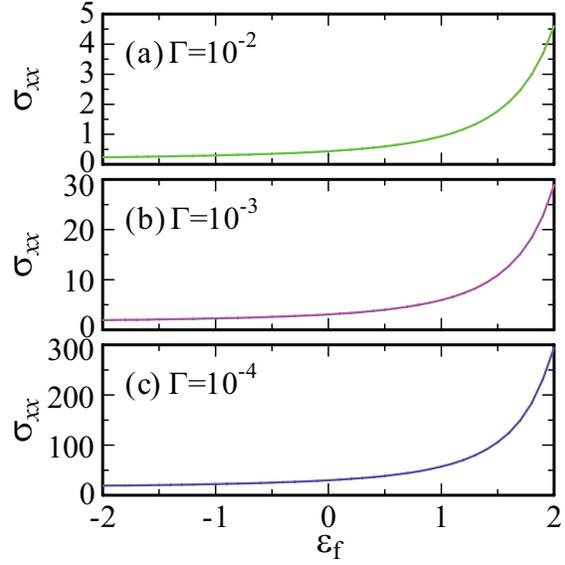}
\caption{ (Color online) 
The $\varepsilon_{\rm f}$ dependence of the conductivity $\sigma_{xx}$ 
for $t=1$, $V=0.3$, and $U=\infty$ 
at $\bar{n}=7/4$ 
with (a) $\Gamma=10^{-2}$, (b) $10^{-3}$, and (c) $10^{-4}$ 
calculated in the $N=1200\times 1200$ lattice sites. 
}
\label{fig:Gm_dep_sxx_Ef}
\end{figure}

Figure~\ref {fig:Gm_dep_sxx_Ef} shows the $\varepsilon_{\rm f}$ dependence of the conductivity $\sigma_{xx}$ for $t=1$, $V=0.3$, and $U=\infty$ at $\bar{n}=7/4$ with (a) $\Gamma=10^{-2}$, (b) $10^{-3}$, and (c) $10^{-4}$ calculated in the $N=1200\times 1200$ lattice sites. 
Almost constant behavior in the deep-$\varepsilon_{\rm f}$ regime, i.e., Kondo regime, and sharp increase in the shallow-$\varepsilon_{\rm f}$ regime, i.e., valence-fluctuation regime appears in every case, although absolute value of $\sigma_{xx}$ increases. 
As analyzed in Eq.~(\ref{eq:sxx0_2}), $\sigma_{xx}$ is proportional to $\Gamma^{-1}$, which can be seen by comparing Fig.~\ref{fig:Gm_dep_sxx_Ef}(b) with Fig.~\ref{fig:Gm_dep_sxx_Ef}(c). 
However, the relation $\sigma_{xx}\propto\Gamma^{-1}$ does not seem to hold simply between Fig.~\ref{fig:Gm_dep_sxx_Ef}(a) and Fig.~\ref{fig:Gm_dep_sxx_Ef}(b). This indicates that the case of Fig.~\ref{fig:Gm_dep_sxx_Ef}(a) cannot be regarded as the small-$\Gamma$ regime where the analysis based on Eq.~ (\ref{eq:sxx0_2}) is valid. This point will be more clearly seen when we calculate the Hall coefficient $R_{\rm H}$, which will be discussed in Fig.~\ref{fig:Gm_dep_RH_Ef} below. 

\begin{figure}
\includegraphics[width=7.5cm]{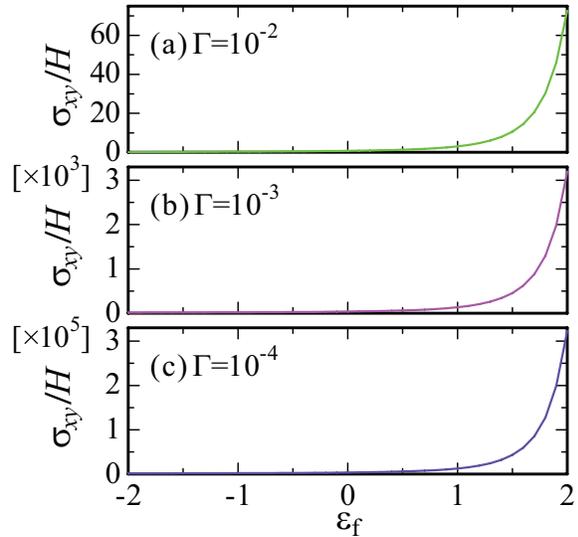}
\caption{ (Color online) 
The $\varepsilon_{\rm f}$ dependence of the Hall conductivity $\sigma_{xy}/H$ 
for $t=1$, $V=0.3$, and $U=\infty$ 
at $\bar{n}=7/4$ 
with (a) $\Gamma=10^{-2}$, (b) $10^{-3}$, and (c) $10^{-4}$ 
calculated in the $N=1200\times 1200$ lattice sites. 
}
\label{fig:Gm_dep_sxyH_Ef}
\end{figure}

Figure~\ref {fig:Gm_dep_sxyH_Ef} shows the $\varepsilon_{\rm f}$ dependence of the Hall conductivity $\sigma_{xy}/H$ for (a) $\Gamma=10^{-2}$, (b) $10^{-3}$, and (c) $10^{-4}$. Almost constant behavior in the Kondo regime and sharp increase in the valence-fluctuation regime appears in every case. 
The relation $\sigma_{xy}/H\propto\Gamma^{-2}$, which is shown in Eq.~(\ref {eq:sxy0_2}), seems to hold between Figs.~\ref{fig:Gm_dep_sxyH_Ef}(b) and \ref{fig:Gm_dep_sxyH_Ef}(c) but not between  Figs.~\ref{fig:Gm_dep_sxyH_Ef}(a) and \ref{fig:Gm_dep_sxyH_Ef}(b). As noted above, this is due to the fact that $\Gamma=10^{-2}$ cannot be regarded as small $\Gamma$. 

Figure~\ref{fig:Gm_dep_RH_Ef} shows the $\varepsilon_{\rm f}$ dependence of the Hall coefficient 
$R_{\rm H}$ for $V=0.3$ at $\bar{n}=7/4$ 
for a series of damping rates due to impurity scattering;  
$\Gamma=10^{-4}$ (dashed line), $\Gamma=10^{-3}$ (solid line), and 
$\Gamma=10^{-2}$ (dash-dotted line), which are calculated in the $N=1200\times 1200$ lattice sites. 
Almost constant $\varepsilon_{\rm f}$ dependence, $R_{\rm H}\approx 1/(\bar{n}_{\rm hole}e)=4/e$,  appears for $\Gamma=10^{-4}$ and $\Gamma=10^{-3}$. 
However, 
for $\Gamma=10^{-2}$, $R_{\rm H}$ shows a visible deviation from the constant behavior. 

\begin{figure}
\includegraphics[width=7.5cm]{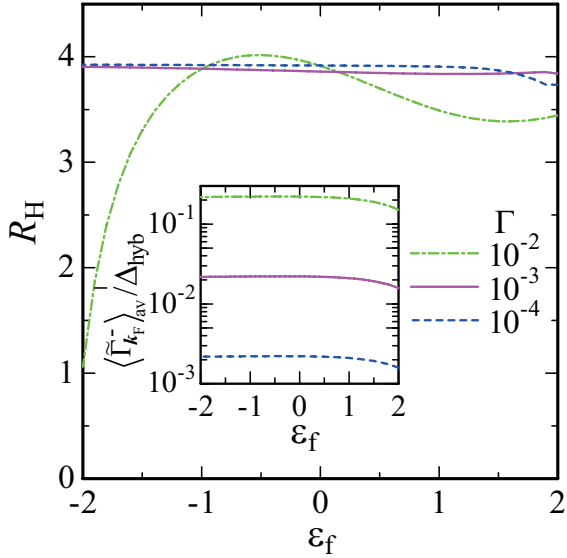}
\caption{ (Color online) 
The $\varepsilon_{\rm f}$ dependence of the Hall coefficient $R_{\rm H}$ 
for $t=1$, $V=0.3$, and $U=\infty$ 
at $\bar{n}=7/4$ 
with $\Gamma=10^{-4}$ (dashed line), $10^{-3}$ (solid line), and $10^{-2}$ (dash-dotted line) 
calculated in the $N=1200\times 1200$ lattice sites. 
Inset: the $\varepsilon_{\rm f}$ dependence of 
$\langle\tilde{\Gamma}_{{\bf k}_{\rm F}}^{-}\rangle_{\rm av}/\Delta_{\rm hyb}$. 
We set $e=1$. 
}
\label{fig:Gm_dep_RH_Ef}
\end{figure}

To quantify the magnitude of the damping rate, in the inset of Fig.~\ref{fig:Gm_dep_RH_Ef} we plot the $\varepsilon_{\rm f}$ dependence of 
the ratio of the damping rate averaged over the Fermi surface 
$\langle\tilde{\Gamma}_{{\bf k}_{\rm F}}^{-}\rangle_{\rm av}$ defined by Eq.~(\ref{eq:Gamma_av}) to $\Delta_{\rm hyb}$ defined by Eq.~(\ref{eq:gap_hyb}). 
These results indicate that 
when the damping rate becomes comparable to about $10~\%$ of the hybridization gap, 
the treatment of the small $\tilde{\Gamma}_{\bf k}^{-}$ 
discussed in Sect.~\ref{sec:small_Gamma} and also using 
Eqs.~(\ref{eq:sxx_2D_Gm0}) and (\ref{eq:sxy_2D_Gm0}) 
are not justified. 
Namely, the contributions from the energies distant from the Fermi energy 
in Eq.~(\ref{eq:sxx_mm}) and Eq.~(\ref{eq:sxy_1}) become relevant to 
$\sigma_{xx}$ and $\sigma_{xy}/H$, respectively. 
For example, the downward deviation of $R_{\rm H}$ with $\Gamma=10^{-2}$ 
seen in the deep-$\varepsilon_{\rm f}$ region in Fig.~\ref{fig:Gm_dep_RH_Ef} 
reflects the tendency that the electron-like curvature of the $\bf k$ points 
in the $\varepsilon<\mu$ region of the lower hybridized band gives 
contributions with 
negative 
sign in $\sigma_{xy}/H$. 

Hence, in the case 
that strong impurity scattering and/or high impurity density 
as well as the extraordinarily-strong correlation gives rise to a large damping rate 
which exceeds $10~\%$ of the hybridization gap,  
$R_{\rm H}$ is not expressed simply 
by the hole density as $1/(\bar{n}_{\rm hole}e)$ even near the half filling at $T=0$. 
It is noted that not only the contributions distant from the Fermi energy to $\sigma_{xx}^{--}$ in Eq.~(\ref{eq:sxx_mm}) and $\sigma_{xy}^{--}$ in Eq.~(\ref{eq:sxy_1}) but also the contributions other than the lower hybridized band are considered to play a significant role in $\sigma_{xx}$ and $\sigma_{xy}$ in such a case. 

\subsubsection{Hybridization dependence 
and pressure dependence in Ce- and Yb-based compounds
}
\label{sec:hyb_dep}

When pressure is applied to the Ce-based compounds, the anions surrounding the Ce$^{+3+\delta}$ ion 
approach the tail of the wavefunction of the 4f electron at the Ce site. 
This causes increase in the crystalline-electronic-field (CEF) level, i.e., $\varepsilon_{\rm f}$ 
increases. 
When pressure is applied to the Yb-based compounds, the negative ions surrounding the 
Yb$^{+3-\delta}$ ion approach, which also makes the 4f-electron level at the Yb site increase. 
Since Yb$^{+3}$ contains 4f$^{13}$ electrons, the hole picture is applied to the periodic Anderson model for the Yb-based systems. Hence, applying pressure makes the 4f-hole level $\varepsilon_{\rm f}$ decrease in Eq.~(\ref{eq:PAM}).   

\begin{figure}
\includegraphics[width=7.5cm]{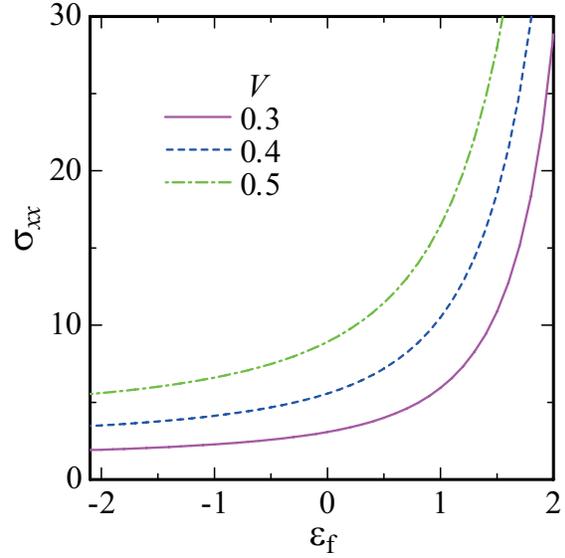}
\caption{
(Color online) The $\varepsilon_{\rm f}$ dependence of the conductivity for $V=0.3$ (solid line), 0.4 (dashed line) and 0.5 (dash-dotted line) at $t=1$, $U=\infty$, $\bar{n}=7/4$ with $\Gamma=10^{-3}$ calculated in the $N=1200\times 1200$ lattice. 
}
\label{fig:sxx_V03_05_Ef}
\end{figure}

In both the Ce- and Yb-based systems, 
the hybridization strength between f and conduction electrons is also expected to increase in general. 
In this subSect., we examine the hybridization dependence of the conductivity, the Hall conductivity, and the Hall coefficient. 

\begin{figure}
\includegraphics[width=7.5cm]{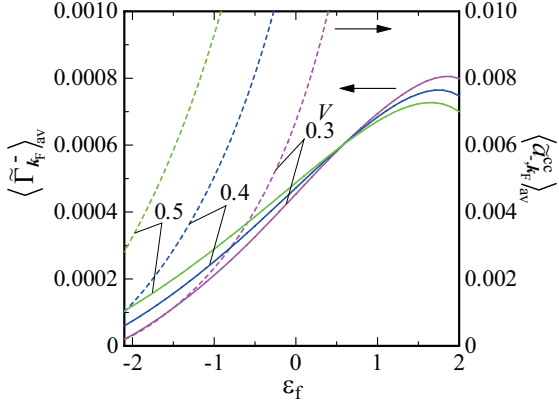}
\caption{
(Color online) The $\varepsilon_{\rm f}$ dependence of the conduction-electron weight factor 
$\langle \tilde{a}_{-,{\bf k}_{\rm F}}^{\rm cc}\rangle_{\rm av}$ 
(dashed line, right axis) and the imaginary part of the selfenergy 
$\langle{\tilde{\Gamma}_{{\bf k}_{\rm F}}^{-}}\rangle_{\rm av}$ (solid line, left axis) for $V=0.3$, 0.4 and 0.5 at $t=1$, $U=\infty$, $\bar{n}=7/4$ with $\Gamma=10^{-3}$ calculated in the $N=1200\times 1200$ lattice. 
}
\label{fig:G_acc_V03_05_Ef}
\end{figure}

In Fig.~\ref{fig:sxx_V03_05_Ef}, we show $\varepsilon_{\rm f}$ dependence of $\sigma_{xx}$ for $V=0.3$ (solid line), 0.4 (dashed line) and 0.5 (dash-dotted line) with $\Gamma=10^{-3}$, which is calculated in the $N=1200\times 1200$ lattice sites.  We see that $\sigma_{xx}$ shifts to larger values as $V$ increases. As analyzed below Eq.~(\ref{eq:sxx0_2}), main contribution to $\sigma_{xx}$ comes from $\tilde{a}^{\rm cc}_{-,{\bf k}}/\tilde{\Gamma}^{-}_{\bf k}$ at the Fermi level. To clarify how hybridization strength affects this quantity, we plot the $\varepsilon_{\rm f}$ dependence of $\langle\tilde{\Gamma}^{-}_{{\bf k}_{\rm F}}\rangle_{\rm av}$ (solid line) and $\langle\tilde{a}^{\rm cc}_{{\bf k}_{\rm F}}\rangle_{\rm av}$ (dashed line) for $V=0.3$, 0.4 and 0.5 in Fig.~\ref{fig:G_acc_V03_05_Ef}. The result shows that the weight factor of conduction electrons $\langle\tilde{a}^{\rm cc}_{{\bf k}_{\rm F}}\rangle_{\rm av}$ shifts to larger values remarkably as $V$ increases while the damping rate of quasiparticles $\langle\tilde{\Gamma}^{-}_{{\bf k}_{\rm F}}\rangle_{\rm av}$ shows no marked enhancement. This can be understood from Eq.~(\ref{eq:damping_MF}). Since the renormalized damping rate is expressed as multiplication of the renormalization factor $z$ and the f-electron weight factor $\tilde{a}^{\rm ff}_{-,{\bf k}}$, as $V$ increases, increase in $z$ and decrease in $\tilde{a}^{\rm ff}_{-,{\bf k}}$ causes cancellation, giving rise to no remarkable enhancement of $\langle\tilde{\Gamma}^{-}_{{\bf k}_{\rm F}}\rangle_{\rm av}$. On the other hand, $\langle\tilde{a}^{\rm cc}_{{\bf k}_{\rm F}}\rangle_{\rm av}$ increases as $V$ increases since the weight of conduction electrons at the Fermi level increases by c-f hybridization as understandable from Eq.~(\ref{eq:acc_renm}). Hence, it turns out that hybridization makes $\tilde{a}^{\rm cc}_{-,{\bf k}_{\rm F}}/\tilde{\Gamma}^{-}_{{\bf k}_{\rm F}}$ increase, which results in increase in $\sigma_{xx}$ in Fig.~\ref{fig:sxx_V03_05_Ef}. 

\begin{figure}
\includegraphics[width=7.5cm]{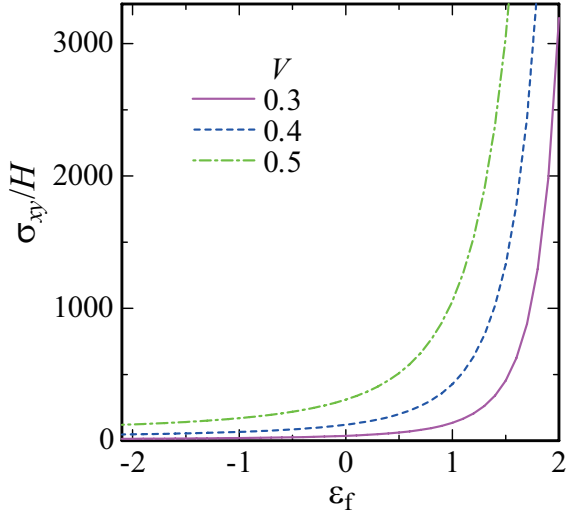}
\caption{
(Color online) The $\varepsilon_{\rm f}$ dependence of the Hall conductivity for $V=0.3$ (solid line), 0.4 (dashed line) and 0.5 (dash-dotted line) at $t=1$, $U=\infty$, $\bar{n}=7/4$ with $\Gamma=10^{-3}$ calculated in the $N=1200\times 1200$ lattice. 
}
\label{fig:sxyH_V03_05_Ef}
\end{figure}

As for the Hall conductivity, the $\varepsilon_{\rm f}$ dependence of $\sigma_{xy}/H$ for $V=0.3$ (solid line), 0.4 (dashed line), and 0.5 (dash-dotted line) with $\Gamma=10^{-3}$ is shown in Fig.~\ref{fig:sxyH_V03_05_Ef}, which is calculated in the $N=1200\times 1200$ lattice sites. As $V$ increases, $\sigma_{xy}/H$ shifts to larger values, similarly to the case of $\sigma_{xx}$. This can be understood from Eq.~(\ref{eq:sxy0_2}). As analyzed below Eq.~(\ref{eq:sxy0_2}), main contribution to $\sigma_{xy}/H$ comes from $(\tilde{a}_{-,{\bf k}}^{\rm cc}/\tilde{\Gamma}_{\bf k}^{-})^2$ at the Fermi level. As shown in Fig.~\ref{fig:G_acc_V03_05_Ef}, $\tilde{a}_{-,{\bf k}_{\rm F}}^{\rm cc}/\tilde{\Gamma}_{{\bf k}_{\rm F}}^{-}$ increases as $V$ increases, which causes increase in $\sigma_{xy}/H$. 

\begin{figure}
\includegraphics[width=7.5cm]{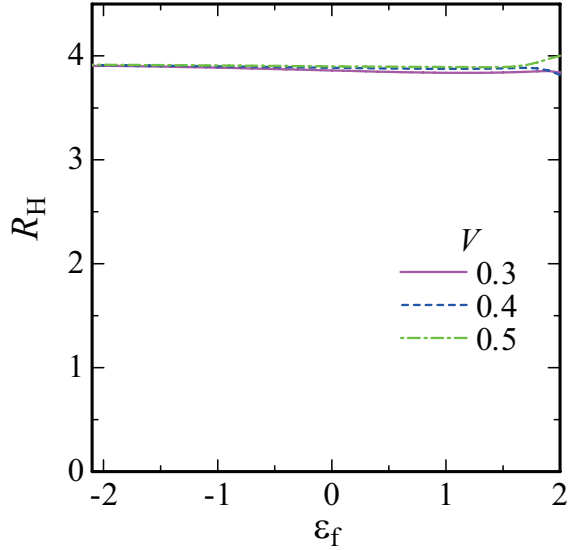}
\caption{(Color online) 
The $\varepsilon_{\rm f}$ dependence of the Hall coefficient $R_{\rm H}$ 
for $t=1$ and $U=\infty$ at $\bar{n}=7/4$ with $\Gamma=10^{-3}$  
calculated in the $N=1200\times 1200$ lattice sites. 
The results 
for $V=0.3$ (solid line), $0.4$ (dashed line), and $0.5$ (dash-dotted line) 
are shown. 
We set $e=1$. 
}
\label{fig:V_dep_RH_Ef}
\end{figure}

In Fig.~\ref{fig:V_dep_RH_Ef}, we plot the $\varepsilon_{\rm f}$ dependence of the Hall coefficient $R_{\rm H}$ 
for a series of hybridization strength; $V=0.3$ (solid line), $V=0.4$ (dashed line), and $V=0.5$ (dash-dotted line) 
with $\Gamma=10^{-3}$, which are calculated in the $N=1200\times 1200$ lattice sites. 
The result shows that even in the cases of $V=0.4$ and $V=0.5$ with increased hybridizations, 
the $\varepsilon_{\rm f}$ dependence of $R_{\rm H}$ remains  
almost the same as that for $V=0.3$.
This is because the factors $\tilde{a}^{\rm cc}_{-,{\bf k}_{\rm F}}/\tilde{\Gamma}^{-}_{{\bf k}_{\rm F}}$ in $\sigma_{xx}$ and $\sigma_{xy}/H$ are canceled out each other in the expression of $R_{\rm H}=\frac{\sigma_{xy}}{H\sigma_{xx}^2}$, as discussed in Sect.~4.2.4. 
This can be 
also
understood from the results shown in Fig.~\ref{fig:Gm_dep_RH_Ef}. 
When $V$ increases, the hybridization gap $\Delta_{\rm hyb}$ increases  
while $\langle\tilde{\Gamma}_{{\bf k}_{\rm F}}^{-}\rangle_{\rm av}$ shows minor change as shown in Fig.~\ref{fig:G_acc_V03_05_Ef}.
Hence, the ratio $\langle\tilde{\Gamma}_{{\bf k}_{\rm F}}^{-}\rangle_{\rm av}/\Delta_{\rm hyb}$ decreases. 
Then, the larger $V$ makes the treatment of the small $\tilde{\Gamma}_{\bf k}^{-}$ works better 
in the calculations of $\sigma_{xx}$, $\sigma_{xy}/H$, and $R_{\rm H}$, 
which reproduces the almost constant $\varepsilon_{\rm f}$ dependence of $R_{\rm H}$.

In the Ce-based compounds, applying pressure makes $\varepsilon_{\rm f}$ and $V$ increase in general. From the results shown in Figs. 4 and 13, and Figs. 7 and 15, $\sigma_{xx}$ and $\sigma_{xy}/H$ show gradual increase in the Kondo (deep-$\varepsilon_{\rm f}$) regime and sharp increase in the valence-fluctuation (shallow-$\varepsilon_{\rm f}$) regime as pressure increases. 
On the other hand, from the results shown in Figs. 8 and 16, almost unchanged $R_{\rm H}$ appears irrespective of the Kondo or valence-fluctuation regime under pressure as far as the system stays in the Fermi liquid. 
Hence, frequently observed behavior in the Ce-based compounds where the residual resistivity decreases gradually in the Kondo regime and drops sharply in the valence-fluctuation regime as pressure increases is naturally explained by the mechanism shown here. 

In the Yb-based compounds, on the other hand, applying pressure makes $V$ increase while $\varepsilon_{\rm f}$ in the hole picture decrease in general. 
Hence, pressure dependence of $\sigma_{xx}$ and $\sigma_{xy}/H$ depends on which factor is more effective. 
In case that the residual resistivity increases sharply in the valence-fluctuation regime and changes to the monotonic increase in the Kondo regime as pressure increases, it indicates that the effect of the $\varepsilon_{\rm f}$ dependence gives major contribution. 
In case that both effects of decreasing $\varepsilon_{\rm f}$ and increasing $V$ are canceled each other, almost unchanged $\sigma_{xx}$ and $\sigma_{xy}/H$ as well as $R_{\rm H}$ are expected to appear under pressure. 

\section{Summary}

We have derived exact formulas for $\sigma_{xx}$ and $\sigma_{xy}$ in the periodic Anderson model 
for $U=0$, which give general expressions of the conductivities in the two-orbital systems with arbitrary band dispersions for $T=0$ as well as finite temperatures. 
On the basis of the theoretical framework for the Fermi liquid based on these formulas, we have studied the ground-state properties of the diagonal and Hall conductivities and the Hall coefficient in the periodic Anderson model with electron correlations on the square lattice,  taking into account the effect of the weak local impurity scattering. The results obtained for the typical case where the Fermi level is located at the lower-hybridized band for the filling of $1\le\bar{n}<2$ with the small damping rate are summarized as follows: 

In the deep-$\varepsilon_{\rm f}$ region where $n_{\rm f}\gsim 0.8$, 
i.e., the Kondo regime, 
almost constant-$\varepsilon_{\rm f}$ dependence of $\sigma_{xx}$ and $\sigma_{xy}/H$ appears as a result of the cancellation of the mass renormalization factors. On the other hand, in the shallow-$\varepsilon_{\rm f}$ region, i.e., the valence-fluctuation regime with $n_{\rm f}\lsim 0.8$, a sharp increase in $\sigma_{xx}$ and $\sigma_{xy}/H$ appears as $\varepsilon_{\rm f}$ increases. This is because the cancellation of the renormalization factors does not occur in the valence-fluctuation regime where the conduction-electron weight factor rapidly increases as $\varepsilon_{\rm f}$ increases while the f-electron damping rate remains to be suppressed. 

On the contrary, the Hall coefficient $R_{\rm H}$ shows an almost constant $\varepsilon_{\rm f}$ dependence in the shallow-$\varepsilon_{\rm f}$ 
region as well as in the deep-$\varepsilon_{\rm f}$ region. 
This is because the renormalization factors expressed as the ratio of the conduction-electron weight factor to the damping rate for f electrons completely cancel out in the expression of $R_{\rm H}$. 
It is shown that $R_{\rm H}$ is expressed as $\frac{1}{\bar{n}_{\rm hole}e}$ with $\bar{n}_{\rm hole}\equiv 2-\bar{n}$ as $\bar{n}$ approaches the half filling, ${\bar n}=2$, while  $R_{\rm H}$ approaches zero as $\bar{n}$ approaches the quarter filling, ${\bar n}=1$. The reason is shown to be naturally understood from the curvatures of the Fermi surface. 

We confirmed that the above conclusions hold at least for the small damping rate for f electrons where it is less than about $10~\%$ of the hybridization gap, which roughly corresponds to the Kondo temperature. It is also confirmed that for the small damping rate the c-f hybridization dependence gives minor effects on the $\varepsilon_{\rm f}$ dependence of $R_{\rm H}$. 

In this paper, we have concentrated on the ground-state properties of the typical periodic Anderson model for the Fermi liquid. 
Theoretically, it has been shown that the magnetically ordered phase generally appears in the deep-$\varepsilon_{\rm f}$ region with $n_{\rm f}$ being close to 1 in Eq.~(\ref{eq:PAM}) if the counter effects such as the magnetic frustration are irrelevant~\cite{WM_CeRhIn5_2010}. In the systems where the inter-orbital Coulomb repulsion between the f electron and the conduction electron which contributes to the energy band located at the Fermi level has a certain magnitude, the quantum critical point (QCP) of the valence transition appears in the ground-state phase diagram~\cite{MW2014,WM2010}. As the magnitude of the c-f hybridization decreases, the QCP of the magnetic transition approaches the QCP of the valence transition and finally coincide each other where the enhanced critical valence fluctuation suppresses the magnetic order giving rise to the first-order magnetic transition~\cite{WM_CeRhIn5_2010,WM2011}. 

When we discuss the transport properties near the QCP of the phase transition such as the magnetic transition and the valence transition, the magnetic fluctuation~\cite{Kontani1999,Kontani2008} and the critical 
valence fluctuation~\cite{MM,MW2014} should be taken into account. 
Indeed, it was shown theoretically that near the QCP of the valence transition in the dirty system, the residual resistivity is enhanced considerably~\cite{MM}, which explains the measurements in the CeCu$_2$Ge$_2$~\cite{Jaccard1999}, CeCu$_2$Si$_2$~\cite{Holmes2004}, 
and CeCu$_2$(Si$_{1-x}$Ge$_{x}$)$_2$~\cite{Yuan2003} systems. 
Hence, when the system approaches the QCP, such effects of the critical fluctuations give rise to additional effects on the results presented in this paper. 

\section*{Acknowledgment}


The authors are grateful to O. Narikiyo and Y. Fuseya for useful discussions on 
the transport theory by showing us their papers prior to publication. 
This work was supported by Grants-in-Aid for Scientific Research (No. 24540378 and No. 25400369) from the Japan Society for the Promotion of Science (JSPS). 
One of us (S.W.) was supported by JASRI (Proposal 
No. 0046 in 2012B, 2013A, 2013B, 2014A, 2014B, and 2015A). 

\appendix
\section{Renormalization factors in conductivity}

In this Appendix, we show that the theoretical framework in Sect.~\ref{sec:sbMF} 
gives essentially the same analytic structure of the conductivity formulated 
on the basis of the Fermi-liquid theory 
in Sect.~\ref{sec:PAM_finiteU}. 
Here we consider the case of  
$\varepsilon_{\bf k}^{\rm f}=\varepsilon_{\rm f}$ and $V_{\bf k}=V$ in Eq.~(\ref{eq:PAM}) 
as discussed in Sect.~\ref{sec:sbMF}. 
Let us start with the first line of Eq.~(\ref{eq:sxx_full}): 
\begin{eqnarray}
\sigma_{xx}^{(1)}&=&
\frac{e^2}{V_0}
\sum_{\bf k}\int_{-\infty}^{\infty}\frac{d\varepsilon}{\pi}
\left(
-\frac{\partial f(\varepsilon)}{\partial\varepsilon}
\right)
\left|G_{\bf k}^{\rm ff \ R}(\varepsilon)\right|^2
v_{{\bf k}x}(\varepsilon)
J_{{\bf k}x}(\varepsilon). 
\label{eq:sxx_A}
\end{eqnarray}
For $\Gamma_{\bf k}^{*}\ll T$,  
$\left|G_{\bf k}^{\rm ff \ R}(\varepsilon)\right|^2$ is evaluated 
as~\cite{Eliashberg}
\begin{eqnarray}
\left|G_{\bf k}^{\rm ff \ R}(\varepsilon)\right|^2
\approx
2\pi i \left(a_{-,{\bf k}}^{\rm ff}\right)^2
\frac{\delta(\varepsilon-E_{\bf k}^{-*})}{i2\Gamma_{\bf k}^{*}}. 
\end{eqnarray}
When the vertex correction in the total current is ignored in Eq.~(\ref{eq:J}), 
the current is given by 
$
J_{{\bf k}x}(\varepsilon)=v_{{\bf k}x}(\varepsilon).  
$
Then Eq.~(\ref{eq:sxx_A}) leads to 
\begin{eqnarray}
\sigma_{xx}^{(1)}=\frac{e^2}{V_0}\sum_{\bf k}
\left(-\frac{\partial f(E_{\bf k}^{-*})}{\partial E_{\bf k}^{-*}}\right)
\left(a_{-,{\bf k}}^{\rm ff}\right)^2
\frac{\left\{v_{{\bf k}x}(E_{\bf k}^{-*})\right\}^2}{\Gamma_{\bf k}^{*}}. 
\label{eq:sxx_A2}
\end{eqnarray}
By using Eq.~(\ref{eq:acc_U}) and Eq.~(\ref{eq:v_def}), 
the relation 
$\left(a_{-,{\bf k}}^{\rm ff}\right)^2
\left\{v_{{\bf k}x}(E_{\bf k}^{-*})\right\}^2
=\left(a_{-,{\bf k}}^{\rm cc}\right)^2
{v_{{\bf k}x}^{\rm c0}}^2
$ 
holds.  
Hence, at sufficiently-low temperatures, $T\approx 0$, Eq.~(\ref{eq:sxx_A2}) 
is expressed as 
\begin{eqnarray}
\sigma_{xx}^{(1)}=\frac{e^2}{V_0}\sum_{\bf k}\delta\left(E_{\bf k}^{-*}-\mu\right)
\frac{\left(a_{-,{\bf k}}^{\rm cc}\right)^2
{v_{{\bf k}x}^{\rm c0}}^2}{\Gamma_{\bf k}^{*}}.
\label{eq:sxx_A_final}
\end{eqnarray}
Since the factor $\left(a_{-,{\bf k}}^{\rm cc}\right)^{-1}$ arises 
from the delta function, $\delta(E_{\bf k}^{-*}-\mu)$ [see Eq.~(\ref{eq:sxx0_2})], 
the factor $a_{-,{\bf k}}^{\rm cc}/\Gamma_{\bf k}^{*}$ finally appears 
in Eq.~(\ref{eq:sxx_A_final}). 
This is the same as Eq.~(\ref{eq:sxx0_2}). 
This confirms the validity of the framework described in Sect.~\ref{sec:sbMF}.

\end{document}